\documentclass[reqno]{amsart}
\usepackage{amscd}

\DeclareMathOperator{\tr}{tr}

\DeclareMathOperator{\hol}{hol}
\DeclareMathOperator{\Map}{Map}
\DeclareMathOperator{\ev}{ev}
\DeclareMathOperator{\Pfaff}{Pfaff}

\DeclareMathOperator{\id}{id}

\theoremstyle{plain}
\newtheorem{theorem}{Theorem}[section]
\newtheorem{lemma}[theorem]{Lemma}
\newtheorem{proposition}[theorem]{Proposition}

\theoremstyle{definition}
\newtheorem{definition}[theorem]{Definition}

\theoremstyle{remark}
\newtheorem{example}{Example}[section]
\newtheorem{note}{Note}[section]

\numberwithin{equation}{section}
\numberwithin{figure}{section}

\newcommand{\cH}{{\mathcal H}}

\newcommand{\cU}{{\mathcal U}}
\newcommand{\cD}{{\mathcal D}}

\newcommand{\C}{{\mathbb C}}
\newcommand{\R}{{\mathbb R}}
\newcommand{\Z}{{\mathbb Z}}

\newcommand{\T}{{\mathbb T}}
\newcommand{\ZZ}{{\mathbb Z}}

\renewcommand{\a}{\alpha}
\renewcommand{\b}{\beta}
\renewcommand{\c}{\gamma}

\renewcommand{\d}{\partial}

\newcommand{\<}{\langle}
\renewcommand{\>}{\rangle}

\begin{document}

\title[Holonomy on D-branes]{Holonomy on D-branes}

\author[A.L. Carey]{Alan L. Carey}
\address[Alan L. Carey]
{Mathematical Sciences Institute\\
Australian National University\\
Canberra ACT 0200 \\
Australia}
\email{acarey@maths.adelaide.edu.au}

\author[S. Johnson]{Stuart Johnson}
\address[Stuart Johnson]
{Pure Mathematics\\
School of Mathematical Sciences\\
University of Adelaide\\
Adelaide, SA 5005 \\
Australia}
\email{sjohnson@maths.adelaide.edu.au}

\author[M.K. Murray]{Michael K. Murray}
\address[Michael K. Murray]
{Pure Mathematics\\
School of Mathematical Sciences\\
University of Adelaide\\
Adelaide, SA 5005 \\
Australia}
\email{mmurray@maths.adelaide.edu.au}

\thanks{The authors acknowledge the support of the Australian
Research Council. ALC completed part of this research for the
Clay Mathematics Institute. SJ acknowledges a University of Adelaide
Scholarship.}


\begin{abstract}This paper shows how to construct anomaly free world
sheet actions in
string theory with $D$-branes. Our method is to use Deligne cohomology and
bundle gerbe theory
to define geometric objects which are naturally associated to
$D$-branes and connections on them. The holonomy of these connections
can be used to cancel global anomalies in the world sheet action.

\end{abstract}
\maketitle

%

%

\section{Introduction}
It has been noted by a number of authors
particularly Freed-Witten and Kapustin \cite{FreWit, Kap} that
the $B$-field, in $D$-brane theory,  defines a Deligne cohomology class and
              this
interpretation has been used to show how anomaly cancellation occurs in the
world sheet action.

The mathematical formalism underlying these observations
starts with a space-time manifold
$M$ with a submanifold $Q\subset M$, the $D$-brane,
and a good open cover $\cU = \{ U_\a \}_{a \in I} $ of $M$
(recall that this means that
every finite intersection of elements in $\cU$ is contractible).
The $B$-field is a collection of smooth  de Rham 2-forms
$\{B_\a\}_{a \in I} $ with ${B_\a}$ defined
on  $U_\a$ and satisfying $dB_\a=dB_\b$ on  $U_\a\cap U_\b$
for all $\a,\b \in I$. Thus $\{dB_\a\}_{\a\in I}$ defines
a 3-form $H$ on $M$. For full treatments see, \cite{DeF},
\cite{Pol},\cite{Wit1}
\cite{Wit2}.

The $B$-field
has various mathematical interpretations
which depend on associated topological and geometric structures.
These interpretations include: a Cech representative
for Deligne cohomology, a differential character, a connection and
curving on a gerbe, a connection and curving on a bundle gerbe, a connection
on a $BS^1$ bundle \cite{Gaj} or some differential geometric structure
on a $PU(H)$ bundle. It is not clear to us if physics
can distinguish
between these different mathematical
realisations of a $B$-field.
In this paper we  focus on the differential character or Deligne
class which is  the minimal
geometric datum necessary to build world sheet actions.

In the simplest case the $B$-field restricts on the $D$-brane $Q$ to the
Stiefel Whitney class of the normal bundle to $Q$.
Then world sheet anomaly cancellation,
or equivalently, the construction of world sheet actions
was investigated in \cite{FreWit}. In this paper we show that for this case
the differential character viewpoint alone suffices.
This refines the results of \cite{FreWit}
in that it eliminates  any dependence of the action on
choices such as open covers and  makes explicit some other necessary
but subtle choices \eqref{eq:pfaff} which affect the
definition of the action.

In order to build a world sheet action in the more difficult situation where
there is a general torsion $B$-field on $Q$ (that is $H=0$ on $Q$)
we need to introduce bundle gerbes and bundle gerbe modules.
These provide an alternative to the Azumaya algebra modules of \cite{Kap}.
Our approach provides a refinement of the conclusions of \cite{Kap}
in making explicit the extent of
dependence on choices made in the construction.

Finally, bundle gerbe modules with infinite dimensional fibre are
needed for world sheet actions in the presence of a
non-torsion $B$-field (i.e $H$ is non-zero on $Q$, see
\cite{BM},\cite{HM}). This case has not been successfully
treated previously.
We propose here a way to produce an anomaly free world sheet action
(this is the main result in the paper).

The paper is organised as follows. Section \ref{sec:action}
contains an overview of our constructions without proofs.
Proofs of the assertions in Section \ref{sec:action}
are presented in the remaining Sections.

We review some Results on Deligne cohomology, its
holonomy and the differential character in Section
\ref{sec:deligne}.
Included here is a discussion of the notion of
transgressing a Deligne two class on a manifold to
a Deligne one class on the loop space of the manifold
although our account emphasises the transgression of the differential
character. This is sufficient to handle the situation considered in 
\cite{FreWit}.

For the more difficult case of anomlay cancellation in the 
presence of general torsion $B$-fields we
need more mathematical structure. This is because the world sheet
action is a priori a section of a non-trivial line bundle.
We use bundle gerbes and bundle gerbe modules to introduce new line
bundles which can be tensored with the original line bundle and
trivialise it. 

Subsection \ref{sec:bg} reviews relevant aspects of
our earlier paper \cite{BCMMS}
where we used bundle gerbes to give a geometric approach to twisted $K$-theory.
Here we explain the geometry of bundle gerbes and
their relation to
Deligne cohomology to connect up with the discussion of Section 3. 
Understanding how the action depends on choices made
in its definition requires us to study gauge transformations of
a bundle gerbe. These generalise the familiar idea of a
gauge transformation on a line bundle.
Then in Subsection \ref{sec:hol_bgm} the holonomy of a connection on
a bundle gerbe module
is introduced motivated by the analogous construction in \cite{Kap}
for Azumaya algebra modules. For torsion $B$ fields,
as in \cite{Kap}, the trace of this bundle gerbe holonomy is
           a section of a line
bundle over the loop space of $Q$ and we tensor this on to our
original world sheet action in order to convert it from a section 
of a non-trivial line bundle to a function.

Our main observation is that a
 modification of this construction may be used
(Subsection \ref{sec:hol_bgm})
to handle the case
of non-torsion $B$-field.

Finally, in Section \ref{sec:cstar},
we round out our account by explaining
the relationship with the approach in \cite{Kap}.
             The point is that in the case of torsion
$B$-field there is a groupoid $C^*$-algebra
with spectrum $M$ which acts on bundle
gerbe modules over $M$.
              This groupoid $C^*$-algebra is continuous trace
              and hence has a Dixmier-Douady  class.
It is then Morita equivalent to any Azumaya algebra with spectrum $M$
having the same Dixmier-Douady class. Azumaya algebras and their
modules are used in \cite {Kap}
to construct twisted $K$-theory.
             It follows then that the $K$-theory of the groupoid algebra
and the Azumaya algebra are the same whenever their Dixmier-Douady
classes are equal
and both give the twisted $K$-theory of $M$. However we do not take
the $C^*$-algebra approach further because we do not know
how to make it work in the non-torsion case.

%

%

\section{Action building}
\label{sec:action}
In this Section we list some basic facts about Deligne cohomology and
show how they can be used to generate anomaly free world sheet actions.
In the subsequent sections we give the mathematical background necessary to
establish these facts.

Let $M$ be a manifold with a submanifold $Q$. Let $\Sigma$ be a
Riemann surface with
a single boundary component which is identified with the circle $S^1$.
Denote by $\Sigma(M)$ the space of all maps
of $\Sigma$ into $M$ and by $L(M)$ the space of all maps of the circle
$S^1$ into $M$. By restricting
a map of $\Sigma $ into $M$ to the boundary circle we obtain a map  of the
circle into $M$. This defines a map we call $\partial \colon
\Sigma(M) \to L(M)$.
            We will be particularly
interested in the subset of maps of $\Sigma$ into $M$ which map the
boundary circle into
the submanifold $Q$. We denote these by $\Sigma_Q(M)$.  There is a
commuting diagram
\begin{equation}
\label{eq:brane}
\begin{array}{ccc}
\Sigma(M) & \stackrel{\partial}{\to } & L(M) \\
               \cup &                              & \cup  \\
\Sigma_Q(M) & \stackrel{\partial}{\to} & L(Q) \\
\end{array}
\end{equation}

World sheet actions are functions on  $\Sigma_Q(M)$. The world sheet actions
that we are interested in  will arise from sections
of line bundles $L \to \Sigma_Q(M)$ constructed from geometric objects
on $Q$ and $M$. The primary geometric object we are interested in
is the Deligne class which is a geometric interpretation of the $B$ field.

Let us review some basic facts about Deligne cohomology.  On a
manifold $X$ there is the group
$H^p(X, \cD^p) $ of Deligne $p$ classes.  For now we need only a few
results about this.

\bigskip

{\bf\noindent Properties of Deligne classes:}

First there is a  homomorphism (see Subsection \ref{sec:holonomy})
$$
c \colon H^p(X, \cD^p)  \to H^{p+1}(X, \Z).
$$
If $\xi$ is a Deligne $p$ class we call $c(\xi)$ its (characteristic) class.

            There is also a homomorphism (Subsection \ref{sec:holonomy})
$$
\iota \colon \Omega^p(X) \to  H^p(X, \cD^p)
$$
            which sends a $p$ form $\rho$ to a Deligne class $\iota(\rho)$ and
we have  $c(\iota(\rho)) = 0$
for any $p$ form $\rho$. The kernel of $\iota$ is $\Omega_c^p(X)$ the
space of all
closed $p$-forms whose integral over any closed submanifold is $2\pi
i$ times an
integer. This discussion is summarized by the  exact sequence of
groups \eqref{eq:complex}
$$
0 \to \Omega^p_c(X) \to \Omega^p(X) \stackrel{\iota}{\to} H^p(X, \cD^p)
\stackrel{c}{\to} H^{p+1}(X, \Z) \to 0.
$$

There is also a map
$$
H^p(X, \cD^p) \to
\Omega^{p+1}(X)
$$
which associates to a Deligne class $\xi$ its
{\em
curvature } $F_\xi$ which is a closed form. The
de Rham class of
$F_\xi$ is the image of $2\pi i c(\xi)$
in real cohomology.

If
$\gamma \colon \Sigma \to  X$ is a map of a  $p$-dimensional
manifold
$\Sigma$ into $X$
and $\xi \in H^p(X, \cD^p)$ is a Deligne
class there is
a {\em holonomy}  $\hol(\xi, \gamma) \in
\C^\times$.

It is known that $H^2(X, \cD^2)$ is the group
of all
isomomorphism classes of line bundles on $X$ with
connection.  In
this case the  connection determines a curvature and a holonomy
which
are the curvature and holonomy of the corresponding Deligne
class.

\medskip{\noindent\bf Transgression:}

Let $\ev \colon S^1 \times
L(X) \to X $ be the evaluation map
and recall that there is a
transgression map
$$
\tau \colon \Omega^{p+1}(X) \to
\Omega^p(L(X))
$$
defined as follows. If $F \in \Omega^{p+1}(X)$ then
$\tau(F)$ is the
result of pulling back $F$ with $\ev$ to
$\Omega(S^1 \times L(X)$ and
then integrating over the circle. There
is an analogous map
$$
\tau \colon H^{p+1}(X, \ZZ) \to H^p(L(X),
\ZZ).
$$

\medskip{\noindent\bf Deligne class of a torsion class:}

       Next we need a result about torsion classes
(Subsection 3.5). Let $\Z_d \subset U(1)$ be the
group of $d$th roots of unity.  Then to any class $\mu \in H^p(X, \Z_d)$
there is a Deligne class $\alpha(\mu) \in H^p(X, \cD^p)$. The class of
$\alpha(\mu)$ is the image of $\mu$ under the Bockstein map
$H^p(X, \Z_d) \to H^{p+1}(X, \Z)$ induced by the short exact sequence
$$
                  \Z  \stackrel{ \times d} {\to}   d\Z   \stackrel{\exp 2\pi
i}{\to} \Z_d .
$$

\medskip{\noindent\bf Line bundle on loop space:}

       For every Deligne class $\xi$ in $H^2(X, \cD^2)$ there is a line bundle
$L_\xi \to L(X)$ over the loop space of $X$
(Subsection 3.6).  This correspondence is
essentially a homomorphism, that is
$L_{\xi + \eta} = L_\xi \otimes L_\eta$, $L_{-\xi} = L^*_\xi$ and
$L_0 = U(1) \times L(X)$.
It is important to note here that these really are equalities in the
sense that there are
canonical isomorphisms in each case.  The Chern class of $L_\xi$ is
the transgression of the
class of $\xi$.

There is also a natural connection on $L_\xi \to
L(X)$
whose curvature is the transgression of $F_\xi$ and
whose
holonomy is determined as follows. If $\gamma \colon S^1 \times
L(X)$ then
the holonomy around $\gamma$ is the holonomy of $\xi$
around $\ev \circ
\id \times \gamma $ where $\id \times \gamma
\colon S^1 \times S^1 \to S^1 \times
L(X) $ is the map $(\id \times
\gamma)(\theta, \phi) = (\theta, \gamma(\phi))$.

\medskip{\noindent\bf Sections of the line bundle on loop space:}

We are interested in sections of the line  $L_\xi \to L(X)$ and its
pullback to $\Sigma(X)$.

The first of these arises because there
is a  canonical non-vanishing section (trivialisation)
$$
\phi_\xi \colon \Sigma(X) \to \partial^{-1}(L_\xi)
$$
defined below in equation \eqref{eq:canonical}.

The second case is when $c(\xi) = 0$. Then the
transgression of $c(\xi)$ is zero and hence $c(L_\xi)= 0$. It follows that
$L_\xi$ is trivial or admits a global non-vanishing section. But now there
is not a canonical section. However if we choose
a $\rho$ with $\iota(\rho) = \xi$ then we can construct a section
$$
\chi_\rho \colon L(X) \to L_\xi.
$$
Notice that, from the exact sequence of groups \eqref{eq:complex} mentioned
      above if we change $\rho$ to $\tau$ with $\iota(\tau) = \xi$ then
$\tau - \rho$ is a
closed two form whose
integral over any two surface is an integral multiple of $2 \pi i$.
The two-form $\tau$ also defines a section $\chi_\tau$ of  $L_\xi$ so we must
have that $\chi_\rho = w \chi_\tau$ for
some function $w \colon L(X) \to U(1)$. The function
$w$ is defined as follows. If $\sigma$ is a map of a disk $D$
into $X$ with boundary a loop $\gamma$
the function
$$
w(\gamma) = \exp(\int_D  \sigma^*(\tau - \rho) )
$$
is well-defined and independent of the choice of $\sigma$. This
construction is, of course, just the definition of the Wess-Zumino-Witten
action of $\tau - \rho$. We will see in Subsections 4.8 and 4.9 how to
understand this fact in terms of gauge transformations of bundle gerbes.

\bigskip

With these observations we can construct world sheet actions.  We start
with the following result
of Freed and Witten \cite{FreWit}. The theory
  of elliptic
operators can be used to construct a line bundle $J_Q \to L(Q)$
  with
a section
$$
\Pfaff \colon \Sigma_Q(M) \to \partial^{-1}(J_Q).
$$
 
Let $w_2 \in H^2(Q, \ZZ_2)$  be the
  second  Steifel-Whitney class of the normal bundle of $Q$. This is a
   torsion class so $\alpha(w_2)$ is a Deligne class in $H^3(Q, \cD^3)$.

  The line bundle $J_Q$ has Chern class the transgression of
$c(\alpha(w_2))$
  and a natural flat connection whose holonomy along
$\gamma \colon S^1
  \to Q$ is given by $(\id \times \gamma)^*(w_2)$.
It follows
  from the discussion above that $L_{\alpha(w_2)}$ with its
natural
  connection is isomorphic to $J_Q$ and has the same holonomy.
Hence
  we can regard $\Pfaff$ as a 
 
\begin{equation}
\label{eq:pfaff}
\Pfaff \colon \Sigma_Q(M) \to \partial^{-1}(L_{\alpha(w_2)}).
\end{equation}
  up to a choice of a constant depending on the $D$-brane $Q$. 

\medskip

\noindent{\bf Case 1}. Assume that the $B$-field or
equivalently the Deligne two class $\xi$ it defines (see example 3.3)
on all of $X$ is such that
\begin{equation}
\label{eq:2.1}
c(\xi_{|Q}) = c(\alpha(w_2)).
\end{equation}

It follows that $L_{\xi_{|Q}}\to L(Q)$ and $L_{\alpha(w_2)}\to L(Q)$
are isomorphic
over $L(Q)$.   By choosing some $\rho\in\Omega^2(Q)$
with $\iota(\rho) = \xi_{|Q}- \alpha(w_2)$ we obtain a non-vanishing section
$\chi_\rho$ of $L_{\xi} \otimes L_{\alpha(w_2)}^*$.
        Finally notice
that $\partial^{-1}(L_{-\xi})\to \Sigma(X)$ has a non-vanishing section
$\phi_{-\xi}$ over $\Sigma(X)$ and this restricts to a non-vanishing
section $\phi_{-\xi}$
over $\Sigma_Q(X)$.  We can now put all the pieces together. The tensor product
\begin{equation}
\label{eq:2.2}
            W(\rho, \xi) =  \Pfaff \otimes \partial^{-1}(\chi_\rho) \otimes
\phi_{-\xi}
\end{equation}
(where  $\partial^{-1}(\chi_\rho)$  denotes the pullback of
the section $\chi_\rho$)
is a section of
$\partial^{-1}(L_{\alpha(w_2)} \otimes L_{\xi_{|Q}} \otimes L_{\alpha(w_2)}^*
\otimes L_{\xi_{|Q}}^*) $ and hence is a   function on $\Sigma_Q(M) $.
In \cite{FreWit} $\chi_\rho$ is regarded as a kind of connection
(it is their $A$-field).
Notice that if we change $\rho$ to $\tau$ subject to
requiring that $\iota(\tau) = \xi_{|Q}- \alpha(w_2)$ and $\sigma \in
\Sigma(M)$ then we have (using $*$ to denote pullback of forms):
\begin{equation}
\label{eq:2.3}
W(\rho, \xi)(\sigma)  = W(\tau, \xi)(\sigma) w(\tau-\rho)(\partial^* \sigma).
\end{equation}

\medskip

\noindent{\bf Case 2.} The $B$-field is torsion on restriction to $Q$ 
but \eqref{eq:2.1} does not
hold, that is, the difference between
$c(\xi_{|Q})$, which comes from the $B$-field,
and  $c(\alpha(w_2))$ is non-zero.

In this case, in order to cancel the anomaly we need an auxiliary
geometric structure. In \cite{Kap} Azumaya algebras played this role.
Here we use bundle gerbes, bundle gerbe modules and connections on
these to give ingredients that we can feed into the world sheet action
to cancel the anomaly which is essentially
\begin{equation}
\label{eq:2.4}
            c(\xi_{|Q}) - c(\alpha(w_2)).
\end{equation}

We will show in Section \ref{sec:bg} that any  bundle gerbe with
connection and curving
gives rise to a
Deligne two class.
            If this Deligne class is torsion the bundle gerbe admits
so-called bundle gerbe modules.  If $A$ is a connection on a bundle
gerbe module
for a bundle gerbe with Deligne class $\eta$ over a manifold $X$ then
we prove in Subsection \ref{sec:hol_bgm}
that the trace of the holonomy of $A$ defines a section $\tr\hol(A)$ of
$L_\eta \to L(X)$. This section
is an extra ingredient that may be used in forming world sheet actions.

Kapustin \cite{Kap} on the other hand
considers a $PU(n)$ bundle $P \to Q$ with
class $\zeta \in H^2(Q, \Z_n)$ and an Azumaya algebra module connection $A$ on
$P \times \C^n$.  As $\zeta$ defines a Deligne
class it defines a line bundle $L_\zeta \to L(Q)$.
The trace of the holonomy of $A$ is a section of this line
bundle and hence pulls back to give a section
$\partial^{-1}(\tr\hol(A))$ of $\partial^{-1}(L_\zeta) \to \Sigma(X)$.
      The product of the Pfaffian of the Dirac operator and
the pull back of the trace of the holonomy of $A$ is now a section of
$\partial^{-1}(L_{\alpha(w_2)} \otimes L_\zeta) $
and to make this trivial Kapustin assumes that $\zeta$ can be chosen
so that
$$
c(\alpha(w_2)) + c(\zeta) = [H]_Q
$$
where $[H] \in H^3(X, \Z)$ is the 3-class arising from the Bockstein
map applied to the $B$-field. It follows that we can
trivialise the line bundle $L_{\alpha(w_2)} \otimes L_\zeta\otimes
L_\xi^*\to L(Q)$
(recall that  $L_\xi$ is the line bundle arising from the
Deligne class $\xi$ or equivalently, from the $B$-field).

Choosing a  $\rho$ with $\iota(\rho)=\xi|_Q -\zeta-\alpha(w_2)$
        we obtain a non-vanishing section $\chi_\rho$ of
$L_\xi\otimes L_\zeta^* \otimes L_{\alpha(w_2)}^*$
We then obtain an action by generalising the construction
\eqref{eq:2.2} to this
situation.

The bundle gerbe version of this is as follows.
Start with the torsion class $\alpha(w_2) - \xi_{|Q}$ on $Q$.
{\it Define} $\zeta$ to be $\alpha(w_2) - \xi_{|Q}$.
There is an associated lifting bundle gerbe
with Dixmier Douady class $c(\zeta)=c(\alpha(w_2))-c(\xi_{|Q})$
(this is described in Subsection \ref{sec:lifting}).
A bundle gerbe module for this lifting bundle gerbe is
just a $PU(n)$ bundle $P\to Q$ for some integer $n$ (Subsection 4.9).
This is the connection with Kapustin's approach and we can proceed
by analogy with \cite{Kap}.
Choose a bundle gerbe module connection $A$ on $P$.
We will show (Subsection \ref{sec:hol_bgm}) that the trace of the holonomy
of $A$ is a section of $L_\zeta\to L(Q)$.

Choosing a stable isomorphism
of $L_{\xi_{|Q}}$ and $L_{\alpha(w_2)} \otimes
L_{\zeta}$ defines a section $\chi$ of
$L^*_{\alpha(w_2)} \otimes
L^*_{\zeta} \otimes L_{\xi_{|Q}}$.
The total world sheet action is then
\begin{equation}
\label{eq:torsion_action}
\Pfaff \otimes \partial^{-1}(\tr\hol(A)) \otimes  \phi_{-\xi} \otimes
\partial^{-1}(\chi).
\end{equation}
Note that in \cite{Kap} the $\chi$ dependence of the action is suppressed.

\medskip

\noindent{\bf Case 3.} The $B$-field is not torsion on restriction to $Q$.

We can proceed as in Case 2 up until we find that the bundle gerbe module for
the lifting bundle gerbe over $Q$
has to have fibre an infinite dimensional Hilbert space $\cH$. Connections
on such a module take their values in the compact operators
on $\cH$ and so cannot have trace class holonomy. Following Section 9
of \cite{BCMMS} we observe that if there are bundle gerbe connections taking
values in the trace class operators on $\cH$ then the difference of the
holonomy of two of these (say $A_1$ and $A_2$) is trace class.
So we fix a reference bundle gerbe module connection $A_1$
taking values in the trace class operators on $\cH$. If $A_2$
is any other trace class operator valued
bundle gerbe connection we will show (Subsection \ref{sec:hol_bgm})
that
$\tr(\hol(A_1)-\hol(A_2))$ is a well defined
section of $L_\zeta  \to  L(Q)$.
Then the world sheet action is the function
\begin{equation}
\label{eq:2.6}
\Pfaff \otimes \partial^{-1}[\tr(\hol(A_1)-\hol(A_2))]
     \otimes  \phi_{-\xi} \otimes
\partial^{-1}(\chi).
\end{equation}

In the remainder of this paper we discuss the mathematics behind all
these constructions. We begin with
the standard description of Deligne cohomology in terms of double
complexes (hyper-cohomology).
We then pass to the description of Deligne cohomology in terms of
differential characters.
This has a number of advantages over the double complex point of
view. In particular it is a
global description not requiring an open cover to be chosen and
moreover it is a
precise description in the sense that the differential character is exactly
the Deligne class rather than a representative of it in some cohomology theory.

%

%

\section{Deligne cohomology}
\label{sec:deligne}

\subsection{Local description}
                  In this Subsection we
review the definition of Deligne cohomology before considering the
anomaly cancellation argument. We let $X$ be a general manifold for
the purposes of this discussion noting that in most cases we will
specialise to $X=Q$.
Recall that for any positive integer $q$ we have the exact sequence
of sheaves $\cD^q$
defined by
\begin{equation}
{\underline {U(1)}} \stackrel{d\log}{\to} \Omega^1 \to \dots \to \Omega^q
\end{equation}
where ${\underline {U(1)}}$ is the sheaf of smooth functions with
values in $U(1)$ and $\Omega^p$ is the sheaf of $p$-forms.
We will define Deligne cohomology in terms of
the sequence $\cD^q$
below and use the notation $H^p(X, \cD^q)$
for these groups although we shall be interested in the special
case $q=p$, that is $H^p(X, \cD^p)$.

Let $\cU = \{ U_\a \}_{a \in I} $ be a good open cover of $X$, that
is every finite intersection of elements on $\cU$ is contractible.
We realise the disjoint union of all the open
sets as
\begin{equation}
\label{eq:nerve}
Y_\cU = \{(x, \a) \mid x \in U_\a \}
\end{equation}
and let $\pi \colon Y_{\cU} \to X$ be the map
$\pi(x, \a) = x$.
The $p$-fold fibre product of $Y_{\cU}$ with itself,
over the map $\pi$ is
\begin{equation}
Y_{\cU}^{[p]} = \{(x, (\alpha_1, \alpha_2, \dots, \alpha_p)) \mid x \in
U_{\alpha_1} \cap
\dots \cap U_{\alpha_p} \} \subset X \times I^p
\end{equation}
which is the disjoint union of all the $p$-fold intersections
$U_{\alpha_1} \cap
\dots \cap U_{\alpha_p}$.
            We define projection maps $\pi_i \colon
Y_{\cU}^{[p]} \to Y_{\cU}^{[p-1]}$
for each $i=1, \dots, p$ by $\pi_i(x, (\a_1, \dots, a_p)) =
(x, (\a_1, \dots, a_{i-1}, a_{i+1}, \dots, a_p))$ and a map
                 $\delta \colon  \Omega^r(Y_{\cU}^{[p]}) \to
\Omega^{r}(Y_{\cU}^{[p-1]})$
by
$$
\delta = \sum_{i=1}^p (-1)^i \pi_i^*.
$$
The space $\Omega^p(Y_{\cU}^{[q]}) $ is the usual space of $p$-form
valued cocycles and the map $\delta$ is the usual coboundary map for
Cech cohomology.  If $\omega \in \Omega^p(Y_{\cU}^{[q]})$ we let
$\omega_{\alpha_1 \dots \alpha_p}$ denote the restriction
of $\omega$ to $U_{\alpha_1} \cap
\dots \cap U_{\alpha_p}$ in the usual way.

To calculate the Deligne cohomology we form the double complex:
\begin{equation}
\label{eq:double_complex}
\begin{array}{ ccccccccc}
                 \ \ \vdots &  & \ \ \vdots & & \ \ \vdots & &&& \ \ \vdots
\\
                 \delta\uparrow &  & \delta\uparrow & & \delta\uparrow &
&&&\delta\uparrow
\\
{\underline {U(1)}}(Y_{\cU}^{[3]}) &  \stackrel{d\log}{\to} &
                 \Omega^1(\Omega^{[3]}) & \stackrel{d }{\to} &
\Omega^2(Y_{\cU}^{[3]}) &
\stackrel{d }{\to}  &\cdots & \stackrel{d}{\to}
                 & \Omega^q(Y_{\cU}^{[3]})
                 \\
                 \delta\uparrow &  & \delta\uparrow & & \delta\uparrow & &
&&\delta\uparrow
\\
{\underline {U(1)}}(Y_{\cU}^{[2]}) &  \stackrel{d\log}{\to} &
                 \Omega^1(Y_{\cU}^{[2]}) & \stackrel{d }{\to} &
\Omega^2(Y_{\cU}^{[2]}) &
\stackrel{d }{\to}  &
\cdots &  \stackrel{d}{\to}&
\Omega^q(Y_{\cU}^{[2]})
\\
\delta\uparrow &  & \delta\uparrow & & \delta\uparrow & && &\delta\uparrow \\
{\underline {U(1)}}(Y_{\cU}) &  \stackrel{d\log}{\to} &
                 \Omega^1(Y_{\cU}) & \stackrel{d }{\to} &  \Omega^2(Y_{\cU}) &
\stackrel{d }{\to}  &
\cdots & \stackrel{d}{\to}&
                 \Omega^q(Y_{\cU})
\end{array}
\end{equation}
The real Deligne cohomology is the cohomology of the double complex
(\eqref{eq:double_complex}) which is calculated by forming the
`diagonal' complex
\begin{equation}
\label{eq:total_complex}
{\underline {U(1)}}(Y_{\cU}) \stackrel{D}{\to} {\underline
{U(1)}}(Y_{\cU}^{[2]})\oplus\Omega^1(Y_{\cU})
\stackrel{D}{\to}{\underline {U(1)}}(Y_{\cU}^{[3]}) \oplus
\Omega^2(Y_{\cU}^{[2]}) \oplus   \Omega^3(Y_{\cU})
\stackrel{D}{\to} \cdots
\end{equation}
where the maps $D$ are defined recursively by (for $g\in {\underline
{U(1)}}(Y_{\cU})$)
\begin{align*}
D(g) &=(\delta(g), d\log g) = (\delta(g), g^{-1}dg) \\
D(g, \omega^1) &=(\delta(g), \delta(\omega^1) - g^{-1}dg , d \omega^1) \\
D(g, \omega^1, \omega^2) &=(\delta(g), \delta(\omega^1) + g^{-1}dg ,
\delta(\omega^2)- d\omega^1, d\omega^2) \\
                 & \quad \vdots
\end{align*}

Standard results in sheaf theory can be applied to show that the
cohomology of the complex \eqref{eq:total_complex} is independent
of the choice of good cover. Similarly we can show that
if $f \colon X \to N$ is a smooth map then we  have a pull-back map
$$
f^* \colon H^p(N, \cD^q) \to  H^p(X, \cD^q)
$$
on Deligne cohomology.

We are interested in the particular case when $p=q$. Then
                 a   Deligne  class is  determined by a collection
$$
(g, \omega^1, \dots, \omega^q) \in {\underline {U(1)}}(Y_{\cU}^{[q+1]})\oplus
\Omega^1(Y_{\cU}^{[q]}) \oplus \
\cdots \oplus \Omega^q(Y_{\cU})
$$
satisfying $D(g, \omega^1, \dots, \omega^q) = 0$ or
$\delta(g) = 1, \delta(\omega^1) = (-1)^{q-1}g^{-1}dg$,
$\delta(\omega^2) = (-1)^{q-2}d\omega^1 ,
\dots, \delta(\omega^q) = d\omega^{q-1}$.
Note that, from its definition as the cohomology of a complex, the Deligne
class of $(g, \omega^1, \dots, \omega^q)$ is unchanged if we replace it
by
\begin{multline}
(g, \omega^1, \dots, \omega^q) + D(h, \mu^1, \dots, \mu^{q-1})=\\ (g\delta(h),
\omega^1 + (-1)^q h^{-1}dh+\delta(\mu^1),
\omega^2 + (-1)^{q-1}d\mu^1 + \delta(\mu^2), \dots,
\omega^q + d\mu^{q-1})
\end{multline}
            where
\begin{equation}
\label{eq:same_class}
(h, \mu^1, \dots, \mu^{q-1}) \in {\underline {U(1)}}(Y_{\cU}^{[q]})
\oplus \Omega^1(Y_{\cU}^{[q-1]})
\oplus \dots
\oplus \Omega^{q-1}(Y_{\cU}).
\end{equation}

Denote by $[g, \omega^1, \dots, \omega^q]$ the Deligne class containing
$(g, \omega^1, \dots, \omega^q)$. Associated to a Deligne class $\xi
= [g, \omega^1, \dots, \omega^q]$
is a $p+1$ form $d\omega^q$.  It is clear from equation
\eqref{eq:same_class} that
this depends only on $\xi$.  Moreover  $\delta(d\omega^q) =d
\delta(\omega^q) = dd\omega^{q-1}$
so that this is a $p+1$ form defined globally on $X$. Denote this
form by $F_\xi$ and
call it the ($q+1$) {\em curvature} of the Deligne class.

\begin{example}
If $p=0$ then a Deligne class is a smooth map $f \colon X \to U(1)$ and
the curvature is the one-form  $f^*(d\theta)$.
\end{example}

\begin{example}
If $p=1$ then a Deligne class $\xi$ can be represented by an isomorphism class
of line bundle with connection.  The curvature of the   Deligne
class is the curvature of the connection.
\end{example}

\begin{example} This is the instance we are mostly concerned with
in this paper.
If $p=2$ then a Deligne class can be represented by a stable isomorphism
class of a bundle gerbe with connection and curving as reviewed in
Subsection \ref{sec:deligne_bg} and originally proved in
\cite{MurSte}.
As explained in \cite{Mur} and reviewed in Subsection \ref{sec:bg_conn}
         a bundle gerbe with connection and curving
gives rise to a three-curvature on the manifold $X$. The
curvature of the Deligne class is precisely this three-curvature.

The $B$-field in string theory may be identified with the
third component ($\omega^2$) of a representative $(g,\omega^1,\omega^2)$
of a Deligne class in $H^2(X, \cD^2)$. The curvature of the Deligne
class is called the $H$-field in string theory.

\end{example}

\subsection{Holonomy of a Deligne class}
\label{sec:holonomy}
Associated to any  Deligne class $$\xi = [g, \omega^1, \dots, \omega^p]$$
is a cohomology class $c(\xi) = [g]$ in $H^{p+1}(X, \Z)$. The image
of $c(\xi)$ in real cohomology is the class of $(1/2\pi i) F_\xi$.
              Let us call the Deligne class
$\xi$ trivial
if the Chern class $c(\xi)$ is zero. Note this is not the same as the
Deligne class being
zero. If $\rho \in \Omega^p(X)$ then we can restrict it to each open set
or equivalently pull it back to $Y_{\cU}$ and hence determine a form
$\pi^*(\rho)$.
This determines a Deligne $p$-class $\iota(\rho )= [1, 0, \dots, 0,
\pi^*(\rho)]$ which is
clearly trivial. Hence we have a sequence of maps
$$
\Omega^p(X) \stackrel{\iota}{\to} H^p(X, \cD^p) \stackrel{c}{\to}
H^{p+1}(X, \Z)
$$
with $c \circ \iota = 0$.  Let $\Omega^p(X)_{(c,0)}$ denote the
subset of $p$-forms which
are closed and whose class in $H^p(X, \R)$ is the image of a class from
$H^p(X, 2\pi i \Z)$. Then there is an exact sequence
\begin{equation}
\label{eq:complex}
0 \to  \Omega^p(X)_{(c,0)} \to \Omega^p(X)\stackrel{\iota}{\to}
H^p(X, \cD^p) \stackrel{c}{\to}
H^{p+1}(X, \Z) \to 0.
\end{equation}

Assume that $X$ is $p$-dimensional so that $H^{p+1}(X, \Z) = 0$.
Then every  Deligne class $\xi$ is trivial so
$\xi = \iota(\rho)$ for some form $\rho$ on $X$ and, assuming that
$X$ is oriented,  we can
define
$$
\hol(\xi, X) = \exp \int_X \rho.
$$
If we choose another $\rho'$ with $\iota(\rho') = \xi$
then
$$
\int_X(\rho - \rho') \in 2 \pi i \Z
$$
so that $\hol(\xi, X)$ is independent of the choice
of $\rho$.
                 Notice that $F_\xi = d\rho$
so if $X$ is the boundary of a $p+1$-dimensional manifold $Y$ then we have
\begin{align}
\label{eq:holonomy}
\hol(\xi, X) &= \exp  \int_{\d Y}  \rho \\
                                & = \exp  \int_{Y}  F_\xi
\end{align}
where $X = \partial Y$ has the induced orientation.

                 More generally if $X$ is not necessarily $p$-dimensional,
we can consider a map $\gamma \colon \Sigma^p \to X$ where $\Sigma^p$ is
$p$-dimensional and compact and define
$$
\hol(\xi, \gamma) = \hol(\gamma^*(\xi), \Sigma^p).
$$

Similarly if $X^{p+1}$ is a $p$ dimensional oriented manifold with boundary
$\partial X^{p+1} = \Sigma^p$, a $p$ dimensional manifold, and
$\gamma \colon X^{p+1} \to X$ we have
$$
\hol(\xi, \partial \gamma) = \exp( \int_{X^{p+1}} \gamma^*(F_\xi))
$$
where $\partial \gamma \colon \Sigma^p \to X$ is the restriction
of $\gamma$ to the boundary.

\begin{example}
If $p=0$ then a Deligne class is a smooth map $f \colon X \to U(1)$ and
the one-form associated to the class is $f^*(d\theta)$. The holonomy
of the smooth map is over a point $p$ and is just the
evaluation of $f$ at $p$.
\end{example}

\begin{example}
If $p=1$ then a Deligne class can be represented by an isomorphism class
of line bundle with connection.  The holonomy is the classical
holonomy of a connection.
\end{example}

\begin{example}
If $p=2$ then a Deligne class can be represented by a stable isomorphism
class of a bundle gerbe with connection and curving.  The
holonomy is the holonomy of a connection and curving defined in
\cite{Mur}
and reviewed in Subsection \ref{sec:deligne_bg}.
\end{example}

Using  \eqref{eq:holonomy} we can define the gluing property of
holonomy. Let $\Sigma_i$
for $i=1, 2, 3$ be manifolds of dimension $p$ related as follows.
Assume we have open
sets $U_i \subset \Sigma_i$ for $i = 1, 2$ such that $\Sigma_i - U_i
$ and $U_i$ are manifolds with
(common) boundary.  Moreover assume we have an orientation reversing
diffeomorphism $\phi \colon U_1 \to U_2$  of manifolds
with boundary so that $\partial \phi \colon \partial(\Sigma_1 -
U_1)
\to \partial(\Sigma_2 - U_2)$
is a diffeomorphism. Finally assume that $\Sigma_3$ is the manifold
constructed by using $\partial \phi$
to glue together $\Sigma_1 - U_1$ and $\Sigma_2 -
U_2$.  Consider now
a pair of maps $\gamma_i \colon
\Sigma_i \to X$ such that ${\gamma_2}|_{U_2} \circ \phi =
{\gamma_1}|_{U_1}$. Then there is
an induced map $f_1 \# f_2 \colon \colon \Sigma_3 \to X$. This map
may not be smooth on the
common boundary of the $\Sigma_i - U_i$ but we can still define its
holonomy. Then we have
\begin{proposition}[Holonomy gluing property]
\label{prop:glueing}
In the situation above
$$
\hol(\xi , \gamma_1 \# \gamma_2) = \hol(\xi, \gamma_1)\hol(\xi , \gamma_2).
$$
\end{proposition}
\begin{proof}
Using the definition of holonomy as an integral it is easy to see that
\begin{align*}
\hol(\xi , \gamma_1 \# \gamma_2) =& \hol(\xi, {\gamma_1}|_{(\Sigma_1
- U_1)})
\hol(\xi , {\gamma_2}|_{(\Sigma_1 - U_2)})\\
=& \hol(\xi, {\gamma_1}|_{(\Sigma_1 - U_1)})\hol(\xi,
{\gamma_1}|_{ U_1})
\hol(\xi, {\gamma_1}|_{U_1})^{-1} \hol(\xi ,
{\gamma_2}|_{(\Sigma_1 - U_2)})\\
=&\hol(\xi, {\gamma_1}|_{(\Sigma_1 - U_1)})\hol(\xi,
{\gamma_1}|_{U_1})
\hol(\xi, {\gamma_2}|_{U_2}) \hol(\xi , {\gamma_2}|_{(\Sigma_1 - U_2)})\\
=&\hol(\xi, {\gamma_1}) \hol(\xi , {\gamma_2})
\end{align*}
Here we use the fact that $\phi$ is orientation reversing to deduce that
$$\hol(\xi, {\gamma_1}|_{U_1})^{-1} = \hol(\xi,
{\gamma_2}|_{ U_2}).$$
\end{proof}

A remark may help the reader to visualize the gluing here when $p=2$.
Imagine that
$\Sigma_1$ and $\Sigma_2$ are two balloons that are pressed together
so they touch on an
open disk $U_1 = U_2$.  Cut out the region where the balloons meet
and we obtain the
surface $\Sigma_3$. We are suppressing mention here of the inclusion
maps $f_1$ and $f_2$ of the surfaces into $X = \R^3$.
Notice that it would be easier to state the \ref{prop:glueing}
 as the holonomy of
$U_1$ times the holonomy over $\Sigma_1 - U_1$ equals the holonomy
over $\Sigma_1$ but we cannot
as holonomy is only defined for closed surfaces.

\subsection{Local formulae}
To compare with the calculations in \cite{FreWit} it is useful to have a
local formulation of the holonomy.  We will restrict
attention to a Deligne two class although a general formula is possible.
Formulae of this type have appeared previously in the
work of
Gawedzki \cite{Gaw, GawRei}, Brylinski \cite{Bry} and \cite{Kap}
Kapustin and for Deligne classes of arbitrary
degree in \cite{GomTer}.  In these applications the formulae were
used to define
the holonomy, here we have an intrinsic definition and we will derive
the local formula.
The case of a Deligne class of arbitrary degree is
in \cite{Stu}.

Consider then a Deligne two-class $\xi = [g, k, B]$ relative to an open
cover $\{U_\a\}$ of $X$.   We
               pull this class back to a surface $\Sigma$
without boundary via a map $\sigma \colon
\Sigma \to X$ and
obtain the class $\sigma^*(\xi) = [\sigma^*(g), \sigma^*(k),
\sigma^*(B)]$ relative to
the open cover $\{ \sigma^{-1}(U_\a)\}$ of $\Sigma$. As $\Sigma$ is
two-dimensional this
class is trivial and we have
$$
\sigma^*(g_{\a\b\c}) = h_{\b\c}h^{-1}_{\a\c} h_{\a\b}
$$
and we can find $m_\a$ such that
$$
\sigma^*(k_{\a\b}) + h^{-1}_{\a\b}d h_{\a\b} = m_\b - m_\a.
$$
If follows that $\sigma^*(B)_\a^{-1}- d m_\a$ is a globally defined two-form
the exponential of whose integral over $\Sigma$ is the holonomy.

Assume now that we have a triangulation of $\Sigma$ into faces, edges
and vertices which is
subordinate to the open cover $\{ \sigma^{-1}(U_\a)\}$. That is the
closure of each face is in (at least one)
open set.  For each face $f$ we choose
a particular open set $\sigma^{-1}(U_\rho(f))$ such that $f \subset
\sigma^{-1}(U_\rho(f))$. Similarly for each
edge $e$ and vertex $v$. Then we have
$$
\hol(\Sigma, \xi) = \prod_{f} \exp \left( \int_f
\sigma^*(B_{\rho(f)}) - dm_{\rho(f)} \right)
$$
where we orient each face with the orientation it inherits from $\Sigma$.
Using Stoke's Theorem this becomes
$$
\hol(\Sigma, \xi) = \prod_{f} \exp \left( \int_f \sigma^*(B_{\rho(f)}) \right)
\prod_{e \subset f} \exp \left(- \int_e  m_{\rho(f)} \right)
$$
where the second product is over all pairs $(e,f)$ consisting of an edge
contained in a face. In the integral the edge is oriented by the
face.  For a pair $e \subset f$ we have
$$
-m_{\rho(f)} = -m_{\rho(e)} + \sigma^*(k_{\rho(f)\rho(e)}) +
h^{-1}_{\rho(f)\rho(e)} dh_{\rho(f)\rho(e)}.
$$
Notice that
$$
\sum_{e \subset f} \int_e -m_{\rho(e)}
$$
vanishes as every edge occurs in exactly two faces and with opposite
orientations.
                We use here the fact that
$\Sigma$ is a manifold without boundary. Hence we have, again using
Stoke's theorem, that
$$
\hol(\Sigma, \xi) = \prod_{f} \exp  \left( \int_f \sigma^*(B_{\rho(f)} )\right)
\prod_{e \subset f} \exp \left( \int_e \sigma^*(k_{\rho(f)\rho(e)}) \right)
\prod_{v \subset e \subset f} h_{\rho(f)\rho(e)}(v).
$$
For a triple $v \subset e \subset f$ we have
$$
               h_{\rho(f)\rho(e)}(v) =   \sigma^*(g_{\rho(f)\rho(e)\rho(v)}) (v)
h_{\rho(f)\rho(v)}(v)
                h^{-1}_{\rho(e)\rho(v)}(v)
$$
and substituting again and observing that the remaining $h$ terms
cancel we obtain
\begin{multline}
\hol(\Sigma, \xi) =\\
\prod_{f} \exp  \left( \int_f \sigma^*(B_{\rho(f)}) \right)
\prod_{e \subset f} \exp \left( \int_e \sigma^*(k_{\rho(f)\rho(e)}) \right)
\prod_{v \subset e \subset f} \sigma^*(g_{\rho(f)\rho(e)\rho(v)}(v)).
\end{multline}

\subsection{Differential characters}

We have seen that we can construct from a Deligne cohomology class
$\xi$ of degree $p$ a
holonomy operation and a curvature form $F_\xi$ which satisfy holonomy
gluing (Proposition \ref{prop:glueing}) and the relation in equation
\eqref{eq:holonomy}.  In an appropriate
sense these two data determine the Deligne cohomology class exactly.  The
appropriate sense is the theory of {\em differential characters}.  A
differential character \cite{CheSim, Bry}  is a
pair $(h, F)$ where $h$ is a homomorphism from $Z_p(X)$, the group of
all smooth, closed $p$ chains (cycles) in $X$,
to $U(1)$ and $F$ is a $p+1$ form. These two are required to be related by
$$
h(\partial \mu) = \exp(\int_\mu F)
$$
for any $p+1$ chain $\mu$ (c.f. \eqref{eq:holonomy}).  The
homomorphism condition on $h$
can be interpreted as the holonomy gluing condition.
The set of all such differential characters is denoted by ${\hat
H}^p(X, U(1))$.

Our construction of holonomy and curvature of a Deligne class has
essentially\footnote{This is not completely true as we have defined
holonomy only over cycles which arise as the images of maps of
triangulated manifolds. We will ignore this issue for the remainder of the
discussion as it does not affect what we are doing.}
defined a map $H^p(X, \cD^p) \to {\hat H}^p(X, U(1))$ and it is a result
of \cite{CheSim} that these two spaces are, in fact,
isomorphic.  In the remainder of this paper we shall work primarily
with differential characters as our representation
for Deligne cohomology. Because of this
isomorphism we can reinterpret various maps we have defined
for Deligne cohomology in terms of differential characters.
First notice that the curvature of a differential
character $\xi = (h, F)$ is, of course, $F$.

Secondly the map
$$
\iota \colon \Omega^p(X) \to {\hat H}^p(X, U(1)).
$$
is defined as follows.
Let $h_\rho \colon Z_p(X)  \to U(1)$ be defined
by $h_\rho(\sigma) = \exp\left(\int_\sigma \rho\right)$ for any
$\rho \in \Omega^p(X)$. This is a homomorphism and
we let  $\iota(\rho) = (h_\rho, d\rho)$.

Thirdly there is an induced map $c \colon {\hat H}^p(X, U(1)) \to H^p(X, \Z)$.
We follow the discussion in \cite{Bry}.  Let $C_p(X)$ be
the group of all chains.  Results from group theory imply that
there is a map $\hat h \colon C_p(X) \to \R$ such that
$h(\sigma) = \exp(\hat h(\sigma))$ for any $\sigma \in C_p(X)$.  Then
$$
\int_\mu F - \hat h(\d\mu) \in 2 \pi i \Z
$$
for any $\mu \in C_{p+1}(X)$. Let $\tau(\mu) = (1/2\pi i )
(\int_\mu F - \hat h(\partial\mu) )
$
and notice that $\partial^*(\tau) = 0$ so that $[\tau]  \in
H^{p+1}(X, \Z)$.
It is straightforward to check that changing the choice of $\hat h$ does
not change the class of $[\tau]$ and we define $c(h,F) = [\tau]$.

Our preference for differential characters is due to their
mathematical simplicity and a belief that they
are generally the  observable quantities  in $D$-brane physics.
However there are many situations where we want to work with
geometric objects which determine a
Deligne class rather than with representatives of the Deligne classes or
of the differential characters themselves.
For any $p$ there are a number of such geometric objects, for example
for $p=2$, the case
of interest in this note, there are gerbes, bundle gerbes, local
gerbes in the sense
of Hitchin, $\Z$ bundle two gerbes  and $BS^1$ bundles in the sense
of Gajer \cite{Gaj}. All of
these, when endowed with appropriate notions of connection and
curvature, determine degree two Deligne classes and differential characters.
While we have a bias towards bundle gerbes (evident later in this
article) the formalism for $D$-branes incorporates additional structure beyond
what
we have described here. It is conceivable that one of these geometric
realisations will be preferable when this additional structure is
taken
into account, however our point here is that since we can formulate
the
discussion in terms of Deligne characters, and all the geometric
realisations lead to these, our account is independent of whatever
geometric realisation is chosen.

It is not clear which physical applications can motivate
a preference for one of these geometric realisations over another.
We will not attempt to comment further on this
question here.

\subsection{Deligne class of a torsion cohomology class}
\label{sec:torsion_deligne}
Recall that the short exact sequence of groups
$$
              \Z \stackrel{\times d}{\to} \Z \to \Z_d
$$
induces the Bockstein map $\beta \colon H^p(X, \Z_d) \to H^{p+1}(X, \Z)$.
We will show that there is a map $\eta \colon H^p(X, \Z_d) \to
H^p(X, \cD^p) $  such that $c \circ \eta = \beta$.

Let  $\kappa \in  H^p(X, \Z_d)$.  Choose a
representative $r \in \kappa$. Then $r $ is a homomorphism from
$C^p(X)$, the group
of all $p$ chains, into $\Z_d$. We can restrict this to obtain a homomorphism
from $Z_p(X)$ into $\Z_d$. If we choose another representative $r'$
of $\kappa$ then $(r-r') = \partial^*(s)$ for some $s \in C_{p-1}(X)$ so
that if $\sigma$ is a closed
$p$ chain then $\< r , \sigma \> - \<r', \sigma \> = \< r -r', \sigma \>
=  \< \partial^*(s), \sigma \> = \< s, \partial(\sigma)\> = 0$.
So we have a well-defined homomorphism $h_\kappa \colon Z_p(X) \to \Z_d
\subset U(1)$.
If $\sigma = \partial (\tau)$ then $h_\kappa(\sigma) = \< r, \partial(\tau) \>
= \< \partial^*(r), \tau \>  = 0$. Hence the pair $(h_\kappa, 0)$ where
$0$ is the zero $(p+1)$-form defines a Deligne cohomology class we denote
by $\eta(\kappa)$.
In terms of Cech representatives relative to an open cover we can
             represent $\kappa$ as the $\Z_d \subset U(1)$ valued cocycle
$\kappa_{i_0, \dots, i_p}$
for which $d \kappa_{i_0, \dots, i_p} = 0$ so that
             $\eta(\kappa)  = (\kappa_{i_0, \dots, i_p}, 0, \dots, 0)$ defines a
Deligne class. It is straightforward
to check that $c(\eta(\kappa) ) = \beta(\kappa)$ the image of
$\kappa$ under the
Bockstein map.

\subsection{Line bundles on loop space.}
\label{sec:transgression}
Let $X$ be a manifold of dimension $m$ and $S$ a compact manifold of
dimension $p$.
Consider the evaluation map
$$
\ev \colon S \times \Map(S,X ) \to X.
$$
If $\rho$ is a differential $r+1$ form on $X$ then we can integrate its
pull-back under $\ev$ to obtain an $r-p+1$ form $\ev_*(\rho)$
on $\Map(S, X)$
called the {\em transgression} of $\rho$. This transgression operation
can be extended to act on differential characters, and hence Deligne
cohomology as follows.  Let $(h , F)$ be a differential
character with $h \colon Z_r(X) \to U(1)$ and $F$ an $r+1$ form.
Clearly we can transgress $F$ to an $r+p-1$ form on $\Map(S, X)$.
Let $\sigma \in Z_{r-p}(S)$ and choose a class $\mu$ representing the
generator of
$H_p(S, \Z) = \Z$. Then $\ev_*(\sigma \times \mu) \in Z_r(X)$ and
we can apply $h$. The result is a map $Z_{r-p}(\Map(S, X)) \to U(1)$
the transgression of $F$. It is, in fact, independent of the choice
of representative $\mu$ and together with the transgression of $F$ satisfies
the conditions for a differential character.  It is also possible
to transgress a Cech representative for a Deligne class but the
result is quite complicated and we refer the reader to \cite{GomTer}
for details.

We concentrate now on the case that $S = S^1$ and hence
$\Map(S, X)$ is the loop space   $L(X)$  of all smooth
maps of the circle
into $X$.  In this case the transgression of a Deligne two class
$\xi$   is a Deligne one class on $L(X)$
and hence it defines an isomorphism class of a line bundle
and connection. We now give a geometric construction of this
line bundle and connection.  For convenience
we assume that $X$ is simply connected and let $D$ be a disk with   $D(X)$  the
space of maps of $D$ into  $X$.
            Let $D(X)^{[2]}$ be pairs of maps $\sigma_1 \colon D \to X$
and $\sigma_2 \colon D \to X$ which agree on the boundary
circle, that is
${\sigma_1}_{| \partial{D} }  = {\sigma_2}_{|\partial D}$.
Each such pair   defines a  map  from the
two sphere  (thought of as the union of two copies of the disk) into $X$.
Denote this map by $\sigma_1 \# \sigma_2$ and orient it by the first factor.

Let $(h, F)$ be the differential character of a Deligne two
class $\xi$.   We define a line bundle
$L_\xi \to L(X)$ whose fibre over a circle
$\gamma \colon S^1 \to X$ is equivalence classes of  pairs $(\sigma, z)$
with $\partial(\sigma) = \gamma$ and $z \in \C$ and equivalence
relation $(\sigma, z) \simeq (\sigma', z') $ if
$$
h(\sigma \# \sigma') z = z'.
$$

This means that a section of $L_\xi$ is a function $s \colon
D(X) \to \C$ such
that
\begin{equation}
\label{eq:transform}
s(\sigma) = h(\sigma \# \sigma') s(\sigma').
\end{equation}
         We think
of this as a transformation rule just as tensor,
spinor and gauge fields satisfy transformation rules for the
Spin, Lorentz and gauge groups.  In the case of sections
of $L_\xi$ there is no group but the philosophy is the
same. Notice that this point of view has the advantage that sections
are actually just functions, albeit on a larger space. In particular
if  $h=1$ then
the section transforms as $s(\sigma) = s(\sigma')$ and hence
defines a function on $L(X)$.

The line bundle $L_\xi \to L(X)$ has a natural connection which we
have no need in this paper to
describe. Note however that if $\gamma  \colon S^1 \to L(X)$ is a loop
then it defines naturally a map $\tilde\gamma \colon S^1 \times S^1
\to X$ and the
holonomy of the connection on $L_\xi$ around $\gamma$  is
$h(\tilde\gamma)$. It can
be shown that the Deligne class of this line bundle with connection is the
transgression of the Deligne two class on $X$.

\subsection{Sections of the line bundle on loop space.}
\label{sec:sections}

In the construction of the world-sheet action we need two basic
sections of $L_\xi \to L(X)$ and its pull-back to $\Sigma(X)$ for a
Deligne two class $\xi$ on $X$.
We first define these and then recall how they are used in Case 1.

Consider first
$$
\phi_\xi \colon \Sigma(X) \to \d^{-1}(L_\xi).
$$
A section of the line bundle $\d^{-1}(L_\xi) \to \Sigma(X)$
    at a point $\nu \in \Sigma(X)$ is a function $s \colon D(X) \times_f
\Sigma(X) \to \C$
satisfying
$s(\sigma, \nu) = h(\sigma \# \sigma') s(\sigma', \nu)$
where elements of $D(X) \times_f \Sigma(X)$ are pairs  $(\sigma, \nu)
\in D(X) \times \Sigma(X)$
such that $\d (\sigma ) = \d (\nu)$ and hence  $\d(\sigma) = \d(\sigma')$.
In particular, that the pullback $\partial^{-1}(L_\xi)$ has a
canonical section defined by
\begin{equation}
\label{eq:canonical}
\phi_\xi(\sigma, \nu) = h(\sigma \# \nu).
\end{equation}
It follows from the holonomy gluing property that
\begin{align*}
\phi_\xi(\sigma, \nu) &= h(\sigma \# \nu) \\
                           &=   h(\sigma \# \sigma') h(\sigma' \# \nu)\\
                 &=h(\sigma \# \sigma')\phi_\xi(\sigma', \nu)
\end{align*}
so that $\phi_\xi(\sigma, \nu)$ is indeed a section of
$\partial^{-1}(L_\xi) \to \Sigma(X)$.

          When we form  an  action out of tensors,
spinors and gauge fields we must combine them so the resulting action
transforms
as a scalar.  So too with world-sheet actions. We must combine
various sections of
various bundles
so that the final action transforms as  $s(\sigma, \nu) = s(\sigma',
\nu)$ and hence defines a function of $\sigma$.

Notice that if $\xi$ and $\xi'$ are two
Deligne classes then $h_\xi h_{\xi'} = h_{\xi + {\xi'}} $. So if
we multiply a section of $L_\xi$ and a section of $L_\xi'$ then
it automatically transforms as a section of $L_{\xi + {\xi'}}$. This
means we have canonical
isomorphisms
$$
         L_\xi \otimes L_{\xi'} \to L_{\xi + {\xi'}}.
$$

The other section used in the construction of
the world sheet action is
$$
\chi_\rho \colon L(X) \to L_\xi.
$$
defined for a $\rho$ with $\iota(\rho) = \xi$. To see
how to define this we note that when $\iota(\rho) = \xi$
the holonomy and curvature of $\xi$ are given by
\begin{align}
h_{\iota(\rho)}(\sigma) &=  \exp\left(\int_\sigma \rho \right)  \\
F_{\iota(\rho)} &=  d\rho.
\end{align}
            Note that $ \exp(\int_\sigma \rho) $ and $d\rho$ are
unchanged if we add an integral, closed form to $\rho$, so
as we expect depend only on $\iota(\rho) = \xi$ not on $\rho$.
The section  $\chi_\rho$ of $L_{\iota(\rho)}$ is defined by
$$
\chi_\rho(\sigma) = \exp\left(\int_D \sigma^*(\rho)\right)
$$
and it is easy to check that this satisfies $\chi_\rho(\sigma) =
h_{\iota(\rho)}(\sigma\#\sigma') \chi_\rho(\sigma')$
as required for a  section of $L_{\iota(\rho)}$.
If we change $\rho$ to $\rho+\mu$ where $\mu $ is a
closed two-form
whose integral over any closed surface is $2\pi i$ times an integer then
\begin{equation}
\label{eq:w}
\chi_{\rho+ \mu}(\sigma) = \exp(\int_D \sigma^*(\mu)) \chi_\rho(\sigma).
\end{equation}

Recall how we apply these constructions to Case 1. We have
the diagram \eqref{eq:brane}
\begin{equation}
\begin{array}{ccc}
\Sigma(M) & \stackrel{\partial}{\to } & L(M) \\
               \cup &                              & \cup  \\
\Sigma_Q(M) & \stackrel{\partial}{\to} & L(Q) \\
\end{array}
\end{equation}
and we want to define a function on $\Sigma_Q(M)$. The ingredients
are a Deligne two class ($B$-field) $\xi $ on $M$ and the
(torsion) Steifel-Whitney class $w_2 \in H^2(Q, \Z_2)$ which
together satisfy \eqref{eq:2.1}
$$
c(\xi_{|Q}) = c(\alpha(w_2)),
$$
and the section
$$
\Pfaff \colon  \Sigma_Q(M) \to     \partial^{-1}(L_{\alpha(w_2)}).
$$
defined in \eqref{eq:pfaff}.

First we apply the above constructions to get  $\phi_{-\xi}
\colon \Sigma(M) \to \d^{-1}(L_{-\xi})$ and restrict this to $\Sigma_Q(M)$
to get (by abuse of notation)
$$
\phi_{-\xi}   \colon \Sigma_Q(M) \to \left.
\d^{-1}(L_{-\xi})\right|_{\Sigma_Q(M)}
$$

Secondly, because $c(\xi_{|Q}-\alpha(w_2))= 0$ we can choose $\rho
\in \Omega^2(Q)$
such that $\iota(\rho) = c(\xi_{|Q}) = c(\alpha(w_2))$. Hence, applying
the discussion above but replacing $M$ by $Q$, we  obtain a section
$ \chi_\rho \colon L(Q) \to L_\xi \otimes L_{\alpha(w_2)}^*$ and hence can
pull this back to obtain a section
$$
\partial^{-1}(\chi_\rho) \colon \Sigma_Q(M) \to \d^{-1}(L_\xi)
    \otimes \d^{-1}(L_{\alpha(w_2)})^*.
$$

Combining these three sections we  see that  \eqref{eq:2.3}
    $$
     W(\rho, \xi) =  \Pfaff \otimes \partial^{-1}(\chi_\rho) \otimes \phi_{-\xi}
$$
transforms in such a way that it is a function on $\Sigma_Q(M)$ which
is the world sheet action.

%

%

\section{A geometric interpretation}
\label{bundle_gerbes}

In this Section we are interested in Cases 2 and 3 of Section \ref{sec:action}
that is, general $B$-fields. We
will use bundle gerbes to give a geometric interpretation of
the Deligne character, transgression and the anomaly
cancellation argument.

\subsection{Bundle gerbes}
\label{sec:bg}
Before defining bundle gerbes over a manifold $X$
recall that if
$\pi \colon Y \to X$ is a submersion
(i.e. onto with onto
differential)
then  $X$ can be covered
by open sets $U_\alpha$ such that there are
  sections $s_\alpha
\colon U_\a \to X$
of $\pi$, that is $\pi\circ s_\a = 1$. A fibration is a submersion
   but not all submersions maps are fibrations. For example we can
use the disjoint union $Y_{\cU}$ of a given cover $\cU$
as defined in \eqref{eq:nerve}. The sections are the maps $s_\a \colon
U_\a \to Y_\cU$ defined by $s_\a(x) = (x, \alpha)$.

Recall that a  bundle gerbe\footnote{Strictly speaking what we are
about to define
should be called a hermitian  bundle gerbe but the extra terminology is
overly burdensome.} over $X$ is a pair $(L, Y)$ where
$\pi \colon Y \to X$ is a submersion and
$L$ is a hermitian line bundle $P \to Y^{[2]}$ with a product, that is,
a hermitian  isomorphism
$$
L_{(y_1, y_2)} \otimes L_{(y_2, y_3)} \to L_{(y_1, y_3)}
$$
for every $(y_1, y_2)$ and $(y_2, y_3)$ in $Y^{[2]}$.
We require the product to be smooth in $y_1$, $y_2$ and
$y_3$ but in the interests of brevity we will not state the various
definitions needed to make this requirement precise, they  can be found in
\cite{Mur}.
The product is required to be
associative whenever triple products are defined. Also in \cite{Mur}
it is shown that the existence of the product and the associativity
imply isomorphisms $L_{(y, y)} \simeq \C$ and $L_{(y_1, y_2)} \simeq
L_{(y_2, y_1)}^*$.

If $(L, Y)$ is a bundle gerbe  we can define a
new bundle gerbe, $(L^*, Y)$, the dual of $(L, Y)$, by taking
the dual of $L$.
Also if  $(L, Y)$ and $(J, Z)$ are two bundle gerbes we can define their
product $(L\otimes J,  Y\times_\pi Z)$ where $Y\times_\pi Z
=\{ (y, z) \colon \pi_Y(y) = \pi_Z(z) \} $ is
the fibre product of $Y$ and $Z$ over their projection maps.

A morphism from a bundle gerbe  $(L, Y)$ to a bundle
gerbe  $(J, Z)$ consists of a pair of maps $(g, f)$ where
                        $f \colon Y \to Z$ is a map commuting with the
projection to $X$ and  $g \colon L \to J$ is a bundle map covering the
induced map $f^{[2]} \colon Y^{[2]} \to Z^{[2]}$ and commuting with the
bundle gerbe products on $J$ and $L$ respectively. If $f$ and $g$ are
isomorphisms
then we call $(g, f)$ a bundle gerbe isomorphism.

If $J$ is a (hermitian) line bundle over $Y$ then we can define a
bundle gerbe $\delta(J)$ by $\delta(J) = {\pi_1^{-1}(J)}\otimes
\pi_2^{-1}(J)^*$, that is $\delta(J)_{(y_1, y_2)} = J_{y_2} \otimes J_{y_1}^*$,
where $\pi_i\,:\,Y^{[2]} \to Y$ is the map which omits the $i$th element.
The bundle gerbe product
is induced by the natural pairing
$$
J_{y_2}\otimes J_{y_1}^*\otimes J_{y_3}\otimes J_{y_2}^* \to
J_{y_3}\otimes J_{y_1}^*.$$

A bundle gerbe which is isomorphic
to a  bundle gerbe of the form $\delta(J)$ is  called {\em trivial}.
A choice of $J$ and  a bundle gerbe  isomorphism $\delta(J) \simeq L$ is called
a {\em trivialisation}.  If $J$ and $K$ are trivialisations
of $P$ then we have natural isomorphisms
$$
J_{y_1}\otimes J_{y_2}^* \simeq K_{y_1}\otimes K_{y_2}^*
$$
and hence
$$
J_{y_1}^*\otimes K_{y_1} \simeq J_{y_2}^*\otimes K_{y_2}
$$
so that the bundle $J\otimes K$ is the pull-back of a hermitian line
bundle on $X$. Moreover if $J$ is a trivialisation and
$L$ is a bundle on $X$ then $J \otimes \pi^{-1}(L)$ is
also a trivialisation.  Hence the set of all trivialisations of
a given bundle gerbe is naturally acted on by the set of all
hermitian line bundles on $X$.

One can think of
bundle gerbes as one stage in a hierarchy of objects with
each type of object having a characteristic class in $H^p(X, \Z)$.
For example if $p=1$ we have maps from $X$ to $U(1)$, the characteristic
class is the pull-back of $dz$.  When $p=2$  we have hermitian line
bundles on $X$ with
characteristic class the Chern class. When $p=3$ we have bundle gerbes and
they have a characteristic class $d(L) = d(L, Y) \in H^3(X, \Z)$,
the Dixmier-Douady class of $(L, Y)$.  The Dixmier-Douady class is the
obstruction to the bundle gerbe being trivial.
It is shown in \cite{Mur} that
\begin{theorem}[\cite{Mur}]
\label{th:trivial}
A bundle gerbe $(L, Y)$ has zero Dixmier-Douady class
precisely when it is trivial.
\end{theorem}

                       From \cite{Mur} we also have
\begin{proposition}[\cite{Mur}]
\label{th:dd}
If $L$ and $J$ are bundle gerbes over $X$ then
\begin{enumerate}
\item $d(L^*) = -d(L)$ and
\item$d(L\otimes J) = d(L)
+ d(J)$.
\end{enumerate}
\end{proposition}

We note finally that bundle gerbes behave nicely under pull-back. If $(L,Y)$
is a bundle gerbe over $X$ and $f \colon N \to X$ then we can pull-back
$Y$ and hence $L$ to form a bundle gerbe $(f^{-1}(L), f^{-1}(Y))$ over
$N$. We have $d(f^{-1}(L), f^{-1}(Y)) = f^*(d(L,Y))$.

\subsection{Torsion bundle gerbes}
\label{sec:bg_tor}
The definitions
of bundle gerbe, triviality and the Dixmier-Douady
class can be immediately generalised with $U(1)$
replaced by any abelian group $A$ except that the
Dixmier-Douady class lives in $H^2(X, A)$.  In particular
we can consider bundle gerbes for any cyclic subgroup
$\Z_d \subset U(1)$. The Dixmier-Douady class then lives
in $H^2(X, \Z_d)$ and we call these torsion bundle gerbes
or $\Z_d$ bundle gerbes.

It is natural to think of a torsion bundle
gerbe as a $\Z_d$ subbundle of the $U(1)$ bundle $L \to Y^{[2]}$
which is stable under multiplication.  The $U(1)$
bundle gerbe has Dixmier-Douady class in $H^3(X, \Z)$
which is the Bockstein of the torsion bundle gerbe class in $H^2(X, \Z_d)$.
Notice that there are two different notions of triviality for torsion
bundle gerbes, the first is the
vanishing of the class in $H^2(X, \Z_d)$ or torsion bundle gerbe
triviality and the
second is the vanishing of the associated $U(1)$ bundle gerbe or the
vanishing of the class in $H^3(X, \Z)$. The former
implies the latter but not vice versa.

Standard results in topology tell us that every class in $H^3(X,
Z)$ which is torsion arises as the Bockstein of a class
in some $\Z_d$.  Hence every bundle gerbe with torsion
Dixmier-Douady class is stably isomorphic to a torsion bundle gerbe.

\subsection{Lifting bundle gerbes}
\label{sec:lifting}
A common example of bundle gerbes is the so-called {\em lifting
bundle gerbe}. Let
\begin{equation}
\label{eq:cent_ext}
U(1) \to \hat G \stackrel{\pi}{\to} G
\end{equation}
be a central extension of Lie groups and let $P \to X$ be a
principal $G$ bundle. Then there is a map $g \colon P^{[2]} \to G$
defined by $p_1 g(p_1, p_2) = p_2$. We can consider the central extension
as a $U(1)$ bundle over $G$ and pull it back by $g$ to a $U(1)$ bundle
over $P^{[2]}$. The fibre over $(p_1, p_2)$ is the set of all $\hat g$ in
$\hat G$ such that $p_1 \pi(\hat g) = p_2$. The product structure
on $\hat G$ defines a bundle gerbe product. The resulting bundle
gerbe is called the lifting bundle gerbe of $P \to X$.

Given the bundle $P \to X$ it is natural to ask if there is a
$\hat G$ bundle $\hat P \to X$ such that $\hat P / U(1)$ is
isomorphic to $P$ as a $G$ bundle.  It is well known that this is
true if and only if a certain class in $H^3(X, \Z)$ vanishes.
It is also easy to show \cite{Mur} that such a lift is possible
if and only if the lifting bundle gerbe is trivial. Moreover the
class of the lifting bundle gerbe is the three class obstructing the
lift.

The examples we need in this paper are  torsion  bundle
gerbes. For these the central extension is of the form
\begin{equation}
\Z_d \to \hat G \stackrel{\pi}{\to} G
\end{equation}
for some cyclic subgroup $\Z_d \subset U(1)$.  In this case the
obstruction to lifting the $G$ bundle to a $\hat G$ bundle
lives in $H^2(X, \Z_d)$ and again corresponds with the
Dixmier-Douady class of the torsion bundle gerbe.

\subsection{Stable isomorphism of bundle gerbes}
For bundle gerbes there is a  notion called {\em stable isomorphism}
which corresponds exactly to two bundle gerbes having the
same Dixmier-Douady class.  To motivate this consider the case
of two hermitian line bundles $L \to X$ and $J \to X$ they are isomorphic
if there is a bijective map $L \to J$ preserving all structure, i.e. the
projections to $X$ and the $U(1)$ action on the fibres.  Such isomorphisms
are exactly the same thing as trivialisations of $L^*\otimes J$.  For the
case of bundle gerbes the latter is the correct notion and we have
\begin{definition}
\label{def:stableiso}
A stable isomorphism between bundle gerbes  $(L, Y)$ and $(J, Z)$
is a trivialisation of $L^*\otimes J$.
\end{definition}

We have from \cite{MurSte}

\begin{proposition}
\label{prop:stable}
A stable isomorphism exists from  $(L, Y)$ to $(J, Z)$
if and only if $d(L) = d(J)$.
\end{proposition}

If a stable isomorphism exists from $(L, Y)$ to $(J, Z)$
we say that $(L, Y)$ and $(J, Z)$ are stably isomorphic.

It follows easily that  stable
isomorphism is an equivalence relation.
It was shown in \cite{Mur} that every class in $H^3(X, \Z)$ is
the Dixmier-Douady class of some bundle gerbe. Hence we can deduce
from Proposition \ref{prop:stable} that
\begin{theorem}
\label{th:stableiso}
The Dixmier-Douady class defines a bijection between
stable isomorphism classes of  bundle gerbes and $H^3(X, \Z)$.
\end{theorem}

It is shown in \cite{MurSte} that a morphism from  $(L, Y)$ to $(J, Z)$
induces a stable isomorphism but the converse is not true.

\subsection{Bundle gerbe connections and curving}
\label{sec:bg_conn}
Let $(L, Y)$ be a bundle gerbe over $Y$. Before defining
connections we need a useful long exact sequence
from \cite{Mur}. Let $Y^{[p]} \to X$ be the $p$th fold
fibre product of $Y$ over the projection map to $X$. That
is $Y^{[p]}$ is the subset of $Y^p $ consisting of pairs $(y_1, \dots, y_p)$
with the property that $\pi(y_1) = \pi(y_2) = \dots = \pi(y_p)$.
There are projection maps $\pi_i \colon Y^{[p]} \to Y^{[p-1]}$
which omit the $i$th component.  We use these to define a map
on differential forms
\begin{equation}
\label{eq:delta}
\delta \colon \Omega^q(Y^{[p-1]}) \to \Omega^q(Y^{[p]})
\end{equation}
by
$$
\delta(\eta) = \sum_{i=1}^{p} (-1)^i \pi^*(\eta).
$$
Note that $\delta$ commutes with exterior derivative.
It is shown in \cite{Mur} that the long sequence
\begin{equation}
\label{eq:longexact}
0 \to \Omega^q(X) \stackrel{\pi^*}{\to} \Omega^q(Y)
\stackrel{\delta}{\to} \Omega^q(Y^{[2]}) \stackrel{\delta}{\to}
\Omega^q(Y^{[3]}) \stackrel{\delta}{\to} \dots
\end{equation}
is exact for every $q$.

         A connection $\nabla$ on $L
\to Y^{[2]}$
is called a  bundle gerbe connection if it commutes with the product structure
on $L$.  To be more precise, over $Y^{[3]}$, the bundle gerbe
multiplication defines a bundle isomorphism
$
m \colon \pi_3^{-1}(L) \otimes \pi_1^{-1}(L) \to \pi_2^{-1}(L)
$.
On the bundle $\pi_3^{-1}(L) \otimes \pi_1^{-1}(L)$ we have
the connection $\pi_3^{-1}(\nabla) \otimes \pi_1^{-1}(\nabla)$ and on
$\pi_2^{-1}(L)$ the connection $\pi_2^{-1}(\nabla)$. We require that
these are equal under the isomorphism $m$.
It can be shown \cite{Mur} that bundle gerbe connections exist. The
curvature of a bundle
gerbe connection $F_\nabla$ satisfies $\delta(F_\nabla)= 0$ where
$\delta$ is defined in \eqref{eq:delta}. Using the
exactness of \eqref{eq:longexact} we see that there is a
(not unique) two-form $f$ on $Y$ satisfying $\delta(f) = F$. A
choice of such an $f$ we call a {\em curving} for the bundle gerbe connection.
In string theory we would refer to $f$
as the $B$-field. We have that $\delta(df) = d\delta(f) = dF = 0$ so,
using exactness again, $df = \pi^*(\omega)$ for some
three-form $\omega$ on $X$.  As $\pi^*(d\omega) = d \pi^*(\omega)
= ddf = 0$ we see that $d\omega = 0$.
The three-form $\omega$ is called the
three-curvature of the bundle gerbe connection and curving.
In string theory it is the $H$-field.  As for line
bundles the three-curvature represents the image, in real cohomology,
of the Dixmier-Douady class.

We can extend the notion of stable isomorphism to bundle gerbes with
connection and
curving by saying that a bundle gerbe $(L, Y)$ with connection
$\nabla$ and curving $f$
is trivial if there is a line bundle $J \to Y$ with connection $\nabla_J$ and
a bundle gerbe isomorphism $\delta(J) = L$ which maps
$\delta(\nabla_J)$ to $\nabla$
and for which $f = F_{\nabla_J}$. Then two bundle gerbes with connection and
curving $(L, Y)$ and $(K, X)$ are stably isomorphic if $(L\otimes
J^*, Y\times_f X)$
is trivial, as a bundle gerbe with connection and curving. Then we have

\begin{theorem}[\cite{MurSte}]
The set of all stable isomorphism classes of bundle gerbes with
connection and curving is equal to the Deligne cohomology $H^3(X, \cD^3)$.
\end{theorem}

\subsection{Deligne cohomology of a bundle gerbe with connection and curving.}
\label{sec:deligne_bg}
An explicit map to Deligne cohomology can be defined as follows.
Let $\{ U_\a \}$
be a good open cover of $X$ admitting local sections $s_\a \colon U_\a \to X$.
We can define  a map $s \colon Y_\cU \to Y$,
commuting with projections to $X$, by $s(\alpha, x) = s_\alpha(x)$.
This induces maps $s^{[p]} \colon Y_\cU^{[p]} \to Y^{[p]}$ which can
be used to pull-back the line bundle $L \to Y^{[p]}$ to
a line bundle $(s^{[2]})^{-1}(L) \to   Y_\cU^{[2]}$. As the
pairwise intersections are contractible we can trivialise
the line bundle by sections $\sigma_{\a\b} $ over each $U_\a \cap
U_\b$. Then we can multiply
$\sigma_{\a\b}$ and $\sigma_{\b\c}$ using the bundle gerbe product.
Over $U_\a \cap U_\b \cap U_\c$
we must have $\sigma_{\a\b}\sigma_{\b\c} = g_{\a\b\c} \sigma_{\b\c}$ for some
function $g_{\a\b\c}$ which is, in fact, a Cech cocycle.  Also define
$k_{\a\b} \in \Omega^1(U_\a\cap U_\b)$ by $\nabla \sigma_{\a\b} =
k_{\a\b} \sigma_{\a\b}$ and $f_\a \in \Omega^2(U_\a)$ by $f_\a =
s_\a^*(f)$.
In string theory this is how the $B$-field is usually
presented as a collection of 2-forms. The
triple $[g_{\a\b\c}, k_{\a\b}, f_\a]$ defines a Deligne cohomology
class. The curvature
of this Deligne class is the three-curvature of the bundle gerbe
connection and curving.

It follows that every bundle gerbe connection and curving defines a
holonomy, that is
an number in $U(1)$ assigned to any surface in $X$.  To define this explicitly
consider a bundle gerbe with connection and curving over a surface $\Sigma$.
Then as $H^3(\Sigma, \Z)= 0$ this is a trivial bundle gerbe with a
trivialisation
$J \to Y$. It can be shown \cite{MurSte} that we can find a
connection $\nabla_J$
on $J$ such that $\delta(\nabla_J) = \nabla$.
We say such a
connection is compatible with the bundle gerbe connection. Then
$\delta(F_{\nabla_J} - f) = 0$
so that $F_{\nabla_J} - f = \mu_J$ for some two-form $\mu_J $ on
$\Sigma$.   Define
$$
\hol(\nabla, f, \Sigma) = \exp(\int_\Sigma \mu_J).
$$
We leave it as an exercise to confirm that this is the same as the
holonomy of the Deligne class constructed as in Sec.
\ref{sec:holonomy} from the
bundle gerbe with connection and curving.  As for the
case of Deligne cohomology we often also compute holonomy
of a map $\sigma \colon \Sigma \to X$ by first pulling the
bundle gerbe with its connection and curving back to $\Sigma$.

As the bundle $L \to Y^{[2]}$ for a torsion bundle gerbe has a
reduction to $\Z_d$ it has a canonical flat connection.
Because the curvature of the flat connection vanishes the zero two
form on $Y$ is a curving. The flat connection and zero curving
provide a canonical choice of connection and curving for any
torsion bundle gerbe. We leave it as an exercise for the reader to
show that the  Deligne cohomology class defined by the flat
connection and zero curving is the canonical
Deligne cohomology class of a class in $H^2(X, \Z_d)$ defined in
Section \ref{sec:torsion_deligne}.

\subsection{Local bundle gerbes}
If $\cU = \{ U_\a \}_{a \in I} $ is an open cover of  $X$ and we
define $Y_{\cU}$ as in \eqref{eq:nerve} a bundle gerbe $(L, Y_\cU)$ is
just a
collection of line bundles $L_{\a\b} \to U_\a \cap U_\b$. This is a
gerbe in the sense of Hitchin and Chatterjee.
If we restrict further and require that the cover be good we can
assume all the $L_{\a\b}$ are trivial.  In that case bundle gerbe
multiplication
must take the form
$$
((\alpha, x), w) \otimes ((\beta, x), z) ) \mapsto ((\gamma, x),
g_{\a\b\c}(x) w z))
$$
for some co-cycle $g_{\a\b\c} \colon U_\a \cap U_\b \cup U_\c \to
U(1)$ and, moreover, a connection and curving
define exactly a representative for a Deligne cohomology class
in the double complex \eqref{eq:double_complex}.

The local description of bundle gerbes follows from these results. Choose a
good cover $\cU$ and local sections $s_\a \colon U_\a \to Y$. Then these
define a map $s \colon Y_{\cU} \to Y$ by $s(\a, x) = s_a(x)$ which is fibre
preserving. We can use this to pull-back the bundle gerbe $(L, Y)$ to a stably
isomorphic bundle gerbe $(s^{-1}(L), Y_\cU)$ and calculate locally.

\subsection{Stable isomorphism and gauge transformations}
In the case of abelian gauge theory we are interested in
$U(1)$ bundles with connection and curving and these determine
a Deligne one class.  If we act on the bundle with a gauge
transformation then the Deligne class is unchanged. The converse
is also true. To see this let $L$ be a bundle with connections
$A_1$ and $A_2$ defining the same Deligne class. Pick a
point $m_0\in X$.  For any other point $m$ choose a path $\gamma$
from $m$ to $m'$ and consider the parallel transports $P_1(\gamma)$
and $P_2(\gamma)$ from $L_{m_0} $ to $L_{m}$. These define an isomorphism
$$
P_2(\gamma)P_1(\gamma)^{-1} \colon L_{m}  \to L_{m}.
$$
If we choose another path $\gamma'$ then we have
$$
P_i(\gamma') = P_i(\gamma) \hol(\gamma\#\gamma', A_i)
$$
but $\hol(\gamma\#\gamma', A_1) = \hol(\gamma\#\gamma', A_2)$
so that
$$
P_2(\gamma)P_1(\gamma)^{-1} = P_2(\gamma')P_1(\gamma')^{-1}
$$
       and the
result is a gauge transformation $ g \colon L \to L$. Clearly this maps the
parallel transport for $A_1$ to the parallel transport for $A_2$
and hence maps $A_1$ to $A_2$. We conclude that any two connections
with the same Deligne class differ by a gauge transformation.

For bundle gerbes we know that any two bundle gerbes with
the same Deligne class differ by a stable isomorphism with
connection.  Any
two stable isomorphisms differ by a uniquely determined line bundle
       with connection in the sense that if $J \to Y$ and $K \to Y$ are stable
isomorphisms then there is line bundle $L \to X$ such that $J =
\pi^{-1}(L)\otimes K$. In addition $L$ has a connection and the
isomorphism $J = \pi^{-1}(L)\otimes K$ identifies the connection
on $J$ with the product of the pull-back connection on $\pi^{-1}(L)$
and the connection on $K$.    Note that it
is possible to compose stable isomorphisms but  the composition
is not associative \cite{Ste, MurSte}.

In the case of stable isomorphisms from a bundle gerbe $(P, Y)$
to itself
the situation is somewhat simplified as we have a distinguished
stable isomorphism --- the identity. It follows that
every stable isomorphism from $(P, Y)$ to
$(P, Y)$ is determined by a  line
bundle $J$ on $X$ with connection $\nabla$. We conclude that
a gauge transformation of a bundle gerbe $(L, Y)$ with
connection $\nabla$ and curving $f$ is a line bundle $J \to X$ with
connection $D$. Some
calculation shows that  it defines a  stable isomorphism between
$(L, Y)$ with $\nabla$ and $f$ and $(L, Y)$ with $\nabla$ and $f + \pi^*(F_D)$
where $F_D$ is the curvature of the connection $D$ on $J \to X$.
If we take local sections and represent the Deligne class of
$(L, Y)$ with $\nabla$ and $f$ by $(g_{\a\b\c}, a_\a, f_{\a\b})$ then
the stable isomorphism changes it by addition of
$D(k_{\a\b}, A_\a) = (1, 0,  dA_\a)$ where $k_{\a\b}$
are transition functions for $J$ and $A_\a$ are local connection one-forms
for $D$.

\begin{note}
Hitchin has remarked (2001 Arbeitstagung lecture, Max Planck Institute
Bonn) that gauge transformations for gerbes form a category, they are
certainly not a group.
\end{note}

\subsection{Trivial bundle gerbes}
Consider a bundle gerbe with connection and
curving and Deligne class $\xi$. If the Dixmier-Douady class ($c(\xi)$) is
zero then the bundle gerbe is trivial and we can repeat the discussion
in the definition of holonomy in Subsection \ref{sec:deligne_bg} and
find a global trivialisation $J \to Y$ with connection $\nabla_J$.
      The two-form $\mu_J$
is then a two-form on $X$. If we compare with the sequence
\eqref{eq:complex} we
can show that $\iota(\mu_J) = \xi$  the Deligne class of the bundle gerbe.
As in Subsection \ref{sec:sections} we can use $\mu_J$ to define a section
$\chi_{\mu_J}$ of $L_\xi$ over $L(X)$.

If we
change to another trivialisation $J'$ and connection $\nabla'$ then
there is a bundle $K \to X$ with connection $\nabla_K$ such that
$J' = J \otimes \pi^{-1}(K)$, $\nabla_{J'} = \nabla_{J} \otimes
\pi^{-1}(\nabla_K)$
and $\mu_{J'} = \mu_J + F_K$ where $F_K$ is the curvature
of $\nabla_K$. Then we have (c.f. \eqref{eq:w})
$$
\chi_{\mu_{J'}}(\sigma) = \hol(\nabla_K, \partial(\sigma))
\chi_{\mu_{J}}(\sigma).
$$

Notice that the action of a gauge transformation is precisely that of tensoring
the trivialisation $J$ and its connection $\nabla_J$ with the
pull-back of a line
bundle $K \to X$ with connection $\nabla_K$.  It follows that the
change in the action
\eqref{eq:2.3}  arising from changing the trivialisation
$\chi_\rho$ can be regarded as resulting from a gauge transformation
acting on the trivialisation.
To be precise the gauge transformation $(K, \nabla_K)$ acting results in the
value of the action on a world-sheet $\sigma$ being multiplied by
$$
\exp(\int_\Sigma \sigma^*(F_K))
$$
where $F_K $ is the curvature of $\nabla_K$.

\subsection{Bundle gerbe modules}

Let $(L, Y)$ be a bundle gerbe over a manifold $X$ and let $E \to Y$
be a finite rank, hermitian vector bundle. Assume that there is a
hermitian bundle isomorphism
\begin{equation}
\label{eq:bgmod}
\phi \colon L\otimes \pi_1^{-1}E \stackrel{\sim}{\to}
\pi_2^{-1}E
\end{equation}
which is compatible with the bundle gerbe multiplication in the sense
that the two maps
$$
L_{(y_1, y_2)} \otimes (L_{(y_2, y_3)} \otimes E_{y_3} ) \to
L_{(y_1, y_2)} \otimes E_{y_2} \to E_{y_1}
$$
and
$$
(L_{(y_1, y_2)} \otimes L_{(y_2, y_3)} )\otimes E_{y_3} \to
L_{(y_1, y_3)} \otimes E_{y_3} \to E_{y_1}
$$
are the same.  In such a case we call $E$ a  bundle gerbe module and say
that the bundle gerbe acts on $E$.

Notice that if $E$ has rank one then it is a trivialisation
of $L$. Moreover if $E$  has  rank $r$ then $L^r$
acts on $\wedge^r(E)$ and we deduce
\begin{proposition}
\label{prop:torsion}
If $(L, Y)$ has a bundle gerbe module $Y \to E$ of rank $r$ then its
Dixmier-Douady class $d(L)$ satisfies $r d(L) = 0$.
\end{proposition}

A connection $\nabla_E$ is called a bundle gerbe module connection if
            if the bundle gerbe has a connection and the induced connections on
$L\otimes \pi_1^{-1}E$ and $\pi_2^{-1}E$ are equal under the
isomorphism \eqref{eq:bgmod}.

If the bundle gerbe arises as the  lifting
bundle gerbe associated to a principal $G$ bundle $P \to X$ where
there is a central extension
$$
U(1) \to \hat G \to G
$$
it follows from the definition of bundle gerbe module they are the same
thing as bundles $E \to P$ with $\hat G$ action covering the $G$ action
on $P$ and such that the action of $U(1)$ on any fibre $E_p$ over $p \in P$
is scalar multiplication.
For example in the case of
$$
\Z_n \to SU(n) \to PU(n)
$$
the trivial bundle $V \times P$ is a bundle gerbe module whenever
$V$ carries a representation of $SU(n)$.

\subsection{Holonomy of bundle gerbe modules.}
\label{sec:hol_bgm}
We show in this Subsection how,
given a bundle gerbe module over a manifold,
the holonomy of a connection on a  bundle
gerbe module defines a section of the line  bundle
(defined in Subsection \ref{sec:transgression}) over the loop space
of the manifold.
Although the result is general our applications are when the
manifold in question is the submanifold $Q$ as in Sec. \ref{sec:action}.
Our construction is motivated by the construction in \cite{Kap} using
Azumaya algebra module connections.

Consider a bundle gerbe $(R, Y)$  over $Q$ with a connection and
curving defining a torsion Deligne class $\zeta$. The example we need
in Sec. \ref{sec:action} is the lifting bundle gerbe for a $PU(n)$
principal bundle over $Q$.
Let $E \to Y$ be a bundle gerbe module with a bundle gerbe module
connection $A$.
We wish to define a section $\tr\hol(A) $ of $L_\zeta \to L(Q)$
by constructing a function $s_A \colon D(Q) \to \C$ and showing that
it transforms as in \eqref{eq:transform}.

Let $\sigma \colon D \to Q$ be a map of a disk into $Q$ and pull the
bundle gerbe and connection and module back to $D$. Over $D$ the bundle
gerbe is trivial.  Choose a trivialisation $J$ with connection $\nabla_J$
compatible with the bundle gerbe connection and with curvature $F_J$.
Then we have seen in Sec. \ref{sec:deligne_bg} that
$f - F_J = \pi^*(\mu_J)$ for some $\mu_J$ a two-form on $D$.
Note also that $E \otimes J^* $ with connection $A - \nabla_J$
descends to a
bundle $E_J$ on $D$ with connection $D_J$.  We define
$$
s_A \colon D(Q)  \to \C
$$
by
\begin{equation}
\label{eq:4.6}
s_A(\sigma) = \tr\hol(D_J) \exp(\int_D \mu_J)
\end{equation}
where the holonomy is computed over the boundary
of $\sigma$.  We need to check that $s_A$ is independent of the
choice of $J$ and $\nabla_J$.

\begin{lemma}
The function $s_A \colon D(Q) \to \C$  depends only on $A$ not on the
       choice of trivialisation $J$ or connection $\nabla_J$.
\end{lemma}

\begin{proof}

If we change to another trivialisation $J'$ with
connection $\nabla_J'$ then
there is line bundle $K$ on $D$ with connection $\nabla_K$ such that
$J = \pi^{-1}(K)\otimes J'$
and $\nabla_J = \pi^{-1}(\nabla_K) \otimes \nabla_J'$. Similarly
$E_J = E_J' \otimes K$ and $D_J = D_J'\otimes \nabla_K$.  Hence
\begin{align*}
\hol(D_J) &= \hol(D_J'\otimes \nabla_K) \\
                    &= \hol(D_J') \hol(\nabla_K) \\
                    &= \hol(D_J') \exp(-\int_D F_K).
\end{align*}
so that
\begin{align*}
s_A(\sigma) &= \tr\hol(D_J) \exp(\int_D \mu_J)\\
                   &= \tr\hol(D_J') \exp(-\int_D F_K) \exp(\int_D \mu_J) \\
                &= \tr\hol(D_J') \exp(\int_D \mu_J')\\
\end{align*}
and the function $s_A$ is independent of $J$ and $\nabla_J$.
\end{proof}

Next we have that:
\begin{lemma} The function $s_A$ transforms as a section of $L_\zeta$.
\end{lemma}
\begin{proof}
Assume now that we have two maps $\sigma_i \colon D \to Q$ which agree with
$\sigma$ on the boundary.  Trivialise the pull-back by $\sigma_1 \# \sigma_2 $
of $L_\zeta$ over the whole of the two-sphere. Denote this trivialisation
by $J$ and its connection by $\nabla_J$ and use subscripts $i$ to denote the
restrictions to the two hemispheres $D_1$ and $D_2$. Then
\begin{align*}
s(\sigma_1) &= \tr\hol( D_{J_1}) \exp(\int_{D_1} \mu_{J_1}) \\
                        &= \tr\hol(D_{J_2}) \exp(\int_{D_1} \mu_{J_1})\\
                     &=\tr\hol( D_{J_2}) \exp(\int_{D_2}
\mu_{J_2} \exp(\int_{S^2} \mu_J )\\
               &= s(\sigma_2) \exp(\int_{S^2} \mu_J)
\end{align*}
so that $s$ is a section of $L_\xi$. We use here the fact that  $\sigma_1$
and $\sigma_2$ agree on the boundary of $D$ and that $D_{J_1}$ and
$D_{J_2}$ agree
on this common boundary.
\end{proof}

We now define the section $\tr\hol(A) \colon L(Q) \to L_\zeta$ to be that
given by the function $s_A$. This means we have defined all of the
terms in the tensor product
\eqref{eq:torsion_action}. That the result is a function is a
consequence of these definitions.

In the case that the bundle gerbe is not torsion it was shown in \cite{BCMMS}
that twisted $K$-theory could be constructed from
bundle gerbe modules $E\to Y$ whose structure group was reduced to  the
group of unitaries on an infinite dimensional Hilbert space $\cH$
(isomorphic to the fibres of $E$)
which differ from the identity by a compact
operator. If we require a slightly stronger result, that the
bundle gerbe module have a reduction to the group of
unitaries on $\cH$
that differ from the identity by something which is trace-class
then in the  formula \eqref{eq:4.6} the quantity
$\hol(D_J)$ is a unitary differing from the identity
by a trace-class operator. Choose now two
bundle gerbe module connections $A_1$ and $A_2$ on $E$ so we have
       $\hol(D_{1,J})$  and
$\hol(D_{2,J})$ which are unitaries  differing from the identity
by a trace-class operator. Hence we can define
$$
s(\sigma) = \tr (    \hol(D_{1,J}) - \hol(D_{1,2})   )\exp(\int_{D} \mu_{J}).
$$
To see that this is well defined and a section of $L_\zeta$ is a repeat of the
calculation above. We have
$$
\hol(D_{i,J}) = \hol(D_{i,J'})\hol(\nabla_K)
$$
for $i=1, 2$
so that
$$
\hol(D_{1,J})-  \hol(D_{2,J}) = (  \hol(D_{1,J'})-  \hol(D_{2,J'})
)    \hol(\nabla_K)
$$
giving
$$
\tr(   \hol(D_{1,J})-  \hol(D_{2,J})  ) = \tr(  \hol(D_{1,J'})-
\hol(D_{2,J'}) )\hol(\nabla_K)
$$
and the argument goes through as above to define a section
$\tr(\hol(A_1) - \hol(A_2))$
of $L_\zeta$ over $L(Q)$.

We have made sense of all of the terms in the tensor product (2.7) and
by construction it is a well defined function on $\Sigma_Q(M)$

%

%

\section{Torsion bundle gerbes and $C^*$-algebras}
\label{sec:cstar}

In this Section we show how to relate torsion bundle gerbes to
certain continuous trace $C^*$-algebras. In the course of this
discussion we will explain the relation between Kapustin's work
\cite{Kap} and \cite{BCMMS}.

We start with a principal bundle $\pi:Y\to X$ with structure group
$PU(n)$. All torsion elements of $H^3(X, \Z)$ (Cech cohomology) arise
as the Dixmier-Douady class of the lifting bundle
gerbe $P\to Y^{[2]}$ associated to $\pi:Y\to X$ for some choice of $n$.
Fix one torsion class in $H^3(X,Z)$ and let $P$ be the lifting bundle gerbe
associated to this class by the central extension
$$
\Z_n\to U(n)\to PU(n).
$$

We use the theory of locally compact groupoid $C^*$-algebras
as developed by \cite{R}, \cite{MW1}, \cite{MW2}. To this end observe that
$Y^{[2]}$ is the groupoid of a relation on $Y$ namely
we say $y_1\sim y_2$ if $y_1$ and $y_2$ lie in the same fibre of
$\pi:Y\to X$. The set of equivalence classes under this relation is
$X$. In fact $Y^{[2]}$ is a proper groupoid with unit space $Y$
because it is easy to check that it satisfies the requirement
\cite{MW1} that the map
$\pi_0: Y^{[2]}\to Y\times Y$ which regards $Y^{[2]}$
as a subset of the product $Y\times Y$ is a homeomorphism onto a
closed subset of the product space. Note that the maps
$\pi_1$ and $\pi_2$ from $Y^{[2]}\to Y$ are the range and source maps
respectively of this groupoid which has, as its operations,
the product $(y_1,y_2)(y_2,y_3)= (y_1,y_3)$ and inverse
$(y_1,y_2)^{-1}= (y_2,y_1)$. We identify the unit space
$Y$ with the diagonal $\{(y,y)\ | \ y\in Y\}$.

Now we remark that $Y^{[2]}$ is locally compact and admits a Haar
system. We recall construction of the latter. As $Y\to X$ admits
local sections we can use the resulting local trivialisation to
choose for
$(y_1,y_2)\in Y^{[2]}$ a measure $\lambda^{y_1}$ on the  $\{(y_1,y)\
|\ y\in Y, \pi(y)=\pi(y_1)\}\subset Y^{[2]}$.
In fact we may take for $\lambda^{y_1}$, Haar measure on
$PU(n)$ as the measure on
$\{(y_1,y)\ |\ y\in Y, \pi(y)=\pi(y_1)\}$ using the local
trivialisation to identify these spaces. Note that
a set
$\{(y_1,y)\ |\ y\in Y, \pi(y)=\pi(y_1)\}$ may be identified
with $PU(n)$ in many ways depending on which open set of the cover
we choose. However, we
fix one choice for each fibre throughout. This involves a choice from
only finitely many options as our space $X$ is paracompact and the
cover of $X$ is locally finite.
The set of measures $\{\lambda^{y_1}\ |\ y_1\in Y\}$
is easily seen to define a Haar system on $Y^{[2]}$.
We remark that there is one technical condition on a Haar system that
may not be obvious. This is that if
$C_c(Y^{[2]}$ denotes the continuous functions of compact support on
$Y{[2]}$ then we have for all $f\in C_c(Y^{[2]}$
that the map $(y_1,y_2)\to \int f(y_1,y)d\lambda^{y_1}(y_1,y)$
is continuous. After a moments thought one sees that the construction
of our measures via the local trivialisation guarantees this.

We may describe the groupoid structure on $P$ in a number of ways. To
make use of the results of \cite{MW2} we will use the language of
principal
$\T$-groupoids. This means that we will regard $P$ as an extension of
the groupoid $Y^{[2]}$ in the sense of
Definition 2.2 of \cite{K}. To this end we observe that $P/\T\equiv Y^{[2]}$
because $P$ is a $U(1)$ bundle over $Y^{[2]}$. We may define the
range and source maps of $P$ to be
$r,s:P_{(y_1,y_2)} \to Y$ where $r(z)= (y_1,y_1)$
and $s(z)=(y_2,y_2)$ for $z\in P_{(y_1,y_2)}$. The sense in which $P$
is an extension of $Y^{[2]}$ arises from the existence of
a 2-cocycle on $Y^{[2]}$ defined via the multiplication in $P$.
Recall that $P_{(y_1,y_2)}$ consists of those elements
$u$ of $U(n)$ such that $y_1.p(u)=y_2$ where
$p:U(n)\to PU(n)$ is the projection. We will regard
our extension of $PU(n)$ as a set of pairs $(g,t)$ where $g\in PU(n)$ and
$t\in \Z_n$. This can be achieved globally by choosing
a Borel cross-section $c$ of $p$. Note that as $p$ has discrete
fibres we may choose $c$ to be locally constant. The multiplication
in $U(n)$
is then written
$$(g_1,c(g_1))(g_2,c(g_2))
               = (g_1g_2, c(g_1g_2)\omega(g_1,g_2)) \eqno(*)$$
where $\omega$ is a group 2-cocycle on $PU(n)$.
It is now not hard to recognise $P$ as a principal $\T$-groupoid
as described in \cite{MW1}.

The next step is to identify the Dixmier-Douady class of $P$ regarded
as a bundle gerbe.
It is determined by choosing a good cover
$\{U_\a\}$ of $X$ and transition functions
$g_{\a\b}:U_\a\cap U_\b\to Y$ for the bundle $Y\to X$.
Then the Dixmier-Douady class of $P$ is defined by the multiplication
on $P$. We can write this multiplication using the locally constant
cross-section $c$ and (*) as
$$
c(g_{\a\b}(m))c(g_{\b\c}(m))= c(g_{\a\c})\omega(g_{\a\b}(m),g_{\b\c}(m)).
$$
It follows from this that $\omega$ determines the Dixmier-Douady class of $P$
as a bundle gerbe.

Now we need to describe the $C^*$-algebra associated with this
principal $\T$-groupoid $P$. Let $\Gamma_c(P)$
denote the sections of $P\to Y^{[2]}$ which are of compact support.
These may be thought of as functions:
$f:P\to \C$ satisfying $f(z.t) = tf(z)$ for $z\in P$.
There is a multiplication on $\Gamma_c(P)$ given by
$$f\ast g(z_1)= \int f(z_1z_2)g(z_2^{-1}) d\lambda^{s(z_1)}(\dot z_2),$$
where $\dot z_2$ is the image of $z_2$ under $P\to Y^{[2]}$.
The  involution is
$$
f^*(z_1) =\overline{f(z_1^{-1})}.
$$
We denote by $C^*(P,Y^{[2]},\lambda)$ the $C^*$
completion of $\Gamma_c(P)$ following the notation and definitions of
\cite{MW1}.

The conclusion of the main result of \cite{MW1}
is that the principal $\T$-groupoid
$C^*$-algebra $C^*(P,Y^{[2]},\lambda)$ is continuous
               trace with spectrum $X$. The technical assumption
of \cite{MW2} that
$Y\to X$ admits local sections is clearly satisfied
so that we may apply Section 5 of \cite{MW2}. This states that the
Dixmier-Douady
class of $C^*(P,Y^{[2]},\lambda)$,
is the obstruction to $C^*(P,Y^{[2]},\lambda)$ being Morita
equivalent to the $C^*-$algebra of continuous functions
on $X$ which vanish at infinity $C_0(X)$.

We need to verify that the Dixmier-Douady class of $C^*(P,Y^{[2]},\lambda)$
is the same as the Dixmier-Douady class of $P$ as a bundle gerbe.
This is notationally messy and to save space we refer
the reader to pp128 of
               \cite{MW2}. There, in the discussion centring around equations
(5.5) and (5.6), the
Dixmier-Douady class of $C^*(P,Y^{[2]},\lambda)$ is shown to arise
from $\omega$ in essentially the same fashion as does the
Dixmier-Douady class of $P$
as a bundle gerbe.

Using the known fact \cite{RW} that two continuous trace
$C^*$-algebras with the same spectrum and Dixmier-Douady class are Morita
equivalent we conclude that
if we have an Azumaya algebra over $X$ as in \cite{Kap}
with the same Dixmier-Douady class as $P$ then it must be Morita equivalent to
$C^*(P,Y^{[2]},\lambda)$.

We can see that the bundle gerbe module
$E$ for $P$ is a module for $C^*(P,Y^{[2]},\lambda)$.
               Let $f\in \Gamma_c(P)$
and let $\xi$ be a smooth, compactly supported section of $E$.
It is convenient at this point to regard the bundle gerbe
action on $E$ as being given by a map
$\phi(y_1,y_2): E_{y_2}\to E_{y_1}$.
With this choice we can integrate the $P$ action up
a fibre. So if $\xi$ is a section of $E$ we write
$$\tilde \phi(y_1,y_2).\xi(y_1,y_1) = \phi(y_1,y_2)\xi
[(y_2,y_1)(y_1,y_1)(y_1,y_2)]=\phi(y_1,y_2)\xi(y_2,y_2).$$
The reason for the conjugation action of the groupoid
$Y^{[2]}$ on $Y$ is that this is how it acts on the diagonal.
Now we have to integrate this
to get an action of $f\in \Gamma_c(P)$. We
set, for $z\in P_{y_1,y_2}, \dot z = (y_1,y_2)$ and define
$$f.\xi(r(w)) :=
               \int f(z)\phi(\dot z)\xi(s(w))d\lambda^{r(w)}(\dot z)$$
We could go further but desist at this point
for the reason that we do not know how to make the above
discussion work when the Dixmier-Douady class is non-torsion.
Indeed this was the whole reason for introducing bundle gerbe
modules in the first place: they give a realisation
of twisted $K^0$ without the need to introduce
operator algebras.

\end{document}